\documentclass[12pt]{article}
\usepackage{amsmath, amssymb, amsfonts}
\usepackage[margin=1in]{geometry}
\usepackage{makecell}
\usepackage{booktabs}
\usepackage{enumitem}
\usepackage{graphicx}
\usepackage{float}
\usepackage{rotating}
\usepackage{lineno}
\usepackage[colorlinks]{hyperref}

\title{How Mathematical Forms of Chemotherapy and Radiotherapy Bias Model-Optimized Predictions: Implications for Model Selection}
\author{Changin Oh and Kathleen P.\ Wilkie\\
Department of Mathematics, Toronto Metropolitan University, \\ Toronto, CANADA}

\begin{document}

\maketitle

\begin{abstract}
The move towards personalized treatment and digital twins for cancer therapy requires a complete understanding of the mathematical models upon which these optimized simulation-based strategies are formulated. This study investigates the influence of mathematical model selection on the optimization of chemotherapy and radiotherapy protocols. By examining three chemotherapy models (log-kill, Norton-Simon, and Emax), and three radiotherapy models (linear-quadratic, proliferation saturation index, and continuous death-rate), we identify similarities and significant differences in the optimized protocols. We demonstrate how the assumptions built into the model formulations heavily influence optimal treatment dosing and sequencing, potentially leading to contradictory results. Further, we demonstrate how different model forms influence predictions in the adaptive therapy setting. As treatment decisions increasingly rely on simulation-based strategies, unexamined model assumptions can introduce bias, leading to model-dependent recommendations that may not be generalizable.  This study highlights the importance of basing model selection on a full analysis of bias, sensitivity, practical parameter identifiability and/or inferred parameter posteriors, as a part of the uncertainty quantification process, rather than solely relying on information criterion. Understanding how model choice impacts predictions guiding personalized treatment planning with sufficient uncertainty quantification analysis, will lead to more robust and generalizable predictions. 
\end{abstract}


\clearpage

\section{Introduction}

Personalization of cancer treatment and identification of optimal dosing strategies, whether for individuals (digital twins) or populations (virtual patient cohorts), are built from foundational mechanistic mathematical models~\cite{Strobl2023, Mathur2022, Butner2022, Hamis2019, Craig2023, Stahlberg2022}. Mathematical models, when calibrated with experimental or clinical data, can predict tumour growth and treatment response for various therapies, providing a powerful tool to develop and refine treatment protocols, from the stage of drug development to personalized care in the clinic. Additionally, models of cancer treatments can improve our understanding of the dose-response relationship in cells and tissues, potentially leading to advancements in drug design and delivery methods. 

Recent studies highlight the critical role of model selection in mathematical oncology and the influence of each model on predicted outcomes~\cite{Poleszczuk2018, Wilson2023, Kutuva2023, Porthiyas2024}. For example, Poleszczuk et al.~\cite{Poleszczuk2018} explore how different tumour growth models affect the estimation of PSI and the subsequent radiotherapy protocol recommendations. Using a generalized logistic equation, they demonstrate that PSI calculations and treatment decisions are highly sensitive to the choice of growth model. For instance, logistic growth and Gompertzian growth models produce different PSI values for the same dataset, leading to differences in the radiation dosing recommendations. They further demonstrate that hyperfractionated radiotherapy (lower doses delivered more frequently) is beneficial for specific PSI and growth model combinations, such as moderate growth rates with PSI values between 1.5 and 2.2. Their results demonstrate the challenges of using mathematical models to optimize radiation treatment for a heterogeneous patient population. They concluded that radiation fractionation recommendations based on predictions from a single mathematical model should be critically evaluated and used with caution.

In Kutuva et al.~\cite{Kutuva2023}, they examine how different mathematical forms describing tumour response to radiotherapy influence the estimated radiation dose required for tumour control. They compare two approaches: the direct tumour volume reduction model and the carrying capacity reduction model. While the volume reduction model assumes that radiation directly kills tumour cells, the capacity reduction model assumes that radiation damages the tumour microenvironment, reducing the capacity of the environment to support tumour growth. Their findings show that for tumours with slow growth dynamics (high PSI), the volume reduction model predicts the need for higher radiation doses compared to the capacity reduction model predictions. This discrepancy underscores the importance of understanding the biological assumptions embedded in each model, as the choice between a model based on volume or carrying capacity reduction can significantly affect the predicted optimal treatment regime.

Additionally, Kohandel et al.~\cite{Kohandel2006} explore the sequencing of surgery and chemotherapy for ovarian cancer using mathematical modelling. They consider different tumour growth laws, including Gompertzian and generalized logistic models, combined with various chemotherapy cell-kill hypotheses such as the log-kill (LK), Norton-Simon (NS), and maximum efficacy (Emax) models, to determine the optimal treatment schedule. They show analytically, that for the LK and NS hypothesis, chemotherapy followed by surgery is more effective than surgery followed by chemotherapy when Gompertzian or logistic growth laws are assumed. In the case of exponential growth, however, the two dosing regimes are equal in efficacy. They explore the Emax model computationally, and show that for reasonable parameter values, chemotherapy before surgery performs the best. This study suggests that neoadjuvant chemotherapy, where chemotherapy precedes surgery, may be a preferred treatment regime as it reduces tumour volume prior to surgery, potentially allowing for more effective tumour resection. It also highlights that the model selected can impact the results of the treatment-sequence analysis.

Taken all together, these studies show that the built-in assumptions of mathematical model functional forms can result in different predictions and offer different insights into treatment dosing protocols, emphasizing the importance of careful model selection and critique. In fact, all mathematical models inherently contain some bias, derived from assumptions, hypotheses, and simplifications made during their formulation~\cite{MacLehose2021}. 

In this work, we examine how the assumptions built into various modelling forms of chemo-{} and radiotherapy treatments affect the predicted tumour response and identified optimal protocol.  Understanding the effect of model bias is crucial for analyzing resulting predictions in the context of the bias itself, as bias can lead to over-{} or under-estimation of treatment efficacy. Moreover, the choice of model can influence critical parameters, such as the required total dose needed to achieve tumour control or the optimal dosing protocol, making it essential to critically evaluate the consequences of the underlying assumptions. Specifically, we explore how the mathematical form chosen to describe applied treatment (either chemotherapy or radiotherapy) influences the identified optimal treatment schedule. By doing so, we aim to reveal the implications of model selection on treatment design, and to contribute to a more robust approach to personalized cancer therapy. Consequently, this work emphasizes that model selection is not simply an analysis of information criteria; it is the key to creating treatment plans that patients can trust.

\section{Methods}
Several mathematical models have been developed to describe the effects of chemo-{} and radiotherapy on cancer cells, each capturing unique aspects of the treatment response dynamics. Three standard models of chemotherapy are the log-kill (LK) model, the Norton-Simon (NS) model, and the maximum efficacy (Emax) model~\cite{Kohandel2006}. For radiotherapy, significant models include the gold-standard linear-quadratic (LQ) model~\cite{McMahon2018}(and others), the proliferation saturation index (PSI) model~\cite{Prokopiou2015}, and the continuous death-rate (CDR) model~\cite{Wilson2023}. In this work, we examine these models separately and in combination to determine the underlying bias of each model, and the consequences of this bias on optimized treatment predictions.

\subsection{Mathematical Models for Chemotherapy}
A variety of mathematical models have been developed over the years to describe the dynamics of tumour response to chemotherapy. Early research demonstrated that cell-kill induced by a chemotherapeutic drug is proportional to the tumour population~\cite{Moradi2013}. According to the log-kill (LK) hypothesis, larger tumours are reduced more rapidly than smaller tumours for a given dose. However, subsequent findings revealed that the LK hypothesis does not align with some clinical observations in Hodgkin's disease and leukemia, where larger tumours shrink more slowly than smaller ones~\cite{Simon2006, Norton1976}. As a result, Norton and Simon proposed an alternate hypothesis (NS) wherein chemotherapy leads to a rate of tumour volume regression proportional to the growth rate of an untreated tumour of the same size, suggesting that cell-kill is proportional to the tumour growth rate~\cite{Simon2006}. This hypothesis is generally applicable to many cancer types, particularly those with rapid growth rates~\cite{Schmidt2004}. The maximum efficacy (Emax) hypothesis assumes that cell-kill is proportional to a saturable function of Michaelis-Menten form~\cite{Fister2003}. This hypothesis is based on biological principles where drug effects typically show a sigmoidal response to increasing drug concentration~\cite{Lecca2021, Fernandez2021} or tumour size~\cite{Kohandel2006, Moradi2013, Fister2003, Sharifi2019}. This sigmoidal relationship reflects the saturation of receptors or enzymes involved in the drug mechanism of action. In this study, we take the Emax model that assumes efficacy saturates based on tumour size (rather than drug concentration), as done in~\cite{Kohandel2006}.

The dynamics of these models are defined through a system of ordinary differential equations as:
\begin{align}
    \frac{dV}{dt}&=f(V)-g(C, V) \label{eq:growth} \\
    \frac{dC}{dt}&=\sum_i d \delta(t-t_i)-kC \label{eq:drug}
\end{align}
where $V$ is the tumour volume, $C$ is the drug concentration from applied dose $d$, the delta function $\delta(t-t_i)$ denotes an injection at time $t_i$, and $k$ is the drug clearance rate. Here, $f(V)$ represents the generalized tumour growth rate function and $g(C, V)$ describes the effect of the drug on the tumour (the drug-induced death rate). We consider three cell-kill pharmacodynamic models in this system:
\begin{align}
    g(C, V) = \begin{cases} 
    e_1 CV, & \mbox{LK}\\ 
    e_2 Cf(V), & \mbox{NS}\\ 
    e_3 C\frac{V}{m+V}, & \mbox{Emax}
    \end{cases} \label{eq:chemo}
\end{align}
where each $e_i\, (i=1,2,3)$ represents the drug efficacy for the corresponding model, and $m$ represents the half-maximal sigmoidal shape parameter.

\subsection{Mathematical Models for Radiotherapy}

Several mathematical models have been proposed to predict the response of radiation, including the linear-quadratic (LQ) model~\cite{McMahon2018}, the proliferation saturation index (PSI) model~\cite{Prokopiou2015}, and the continuous death rate (CDR) model~\cite{Wilson2023}. Among these, the most commonly used is the LQ model, which has been widely adopted for its simplicity and effectiveness in describing the biological response of cells to high doses of radiation. Radiation-induced cell death occurs due to permanent double-strand DNA breaks caused directly by ionizing radiation or through interactions between single-strand DNA breaks. The LQ model describes the surviving fraction $SF$ of a cell population following exposure to a dose of $d$ Gy of radiation as
\begin{align}
    SF=e^{-\left(\alpha d+\beta d^2 \right)}
\end{align}
where $\alpha$ Gy\textsuperscript{-1} and $\beta$ Gy\textsuperscript{-2} are the intrinsic radiosensitivity parameters of the irradiated cells~\cite{McMahon2018}. The $\alpha$ term corresponds to single-event cell killing, which occurs when a single ionizing event is sufficient to cause cell death. This linear component reflects the direct damage caused by radiation. In contrast, the $\beta$ term represents multiple-event cell killing, where multiple sub-lethal hits are required to induce cell death. This quadratic component accounts for the accumulation of damage over time and the interactions between multiple DNA breaks. Cells with higher $\alpha$ and $\beta$ are more sensitive to radiation. 

To simulate the effect of radiotherapy, the standard method is to assume an instantaneous reduction in the tumour volume due to cancer cell death. A dose of radiation delivered at time $t_i$ to tumour volume $V_{-}$ (the volume just prior to treatment) is implemented by:
\begin{align}
    V_+=V_{-}SF = V_{-}\left[ e^{-\left( \alpha d+\beta d^2 \right)} \right] \label{eq:LQ}
\end{align}
where $V_{+}$ is the tumour volume immediately after the treatment~\cite{McMahon2018}. This instantaneous volume adjustment is implemented for each radiation fraction in the protocol.

Expanding upon the LQ model, Prokopiou \emph{et al.}~\cite{Prokopiou2015} developed the novel PSI model. PSI ($V/K$), is defined as the ratio of tumour volume ($V$) to estimated carrying capacity ($K$), and it measures the fraction of proliferating cells to the theoretical maximal number of cells due to environmental constraints. The fundamental assumption of the PSI model is that proliferating cells are sensitive to radiation whereas quiescent cells are not. Thus, the proportion of proliferating cells within a tumour significantly influences its response to radiation therapy. Specifically, as tumours approach their carrying capacity, the proportion of proliferating cells decreases, which in turn reduces the overall sensitivity of the tumour volume to radiation.

Similar to the LQ model, the effect of a single dose $d$ of radiation in the PSI model is an instantaneous volume change given by
\begin{align}
    V_+=V_{-} - \left[ 1-e^{-\left( \alpha d+\beta d^2 \right)} \right] V_- \left( 1-\frac{V_-}{K} \right)
    \label{eq:PSI}
\end{align}
at discrete times $t_i$ in the treatment protocol. Here the death fraction is $1-SF = 1-e^{-\left( \alpha d+\beta d^2 \right)}$, following from the LQ model. Radiation-induced cell death is assumed to only affect proliferating cells ($V(1-V/K)$), with quiescent and hypoxic cells assumed to be radioresistant. A higher PSI $(V/K)$ indicates a lower proportion of proliferating cells and hence a treatment-resistant tumour, whereas a lower PSI suggests more proliferating cells and therefore a more radiosensitive tumour.

The LQ and PSI models assume that cell killing happens instantaneously at defined treatment times, which oversimplifies reality. In practice, radiation therapy is delivered through several pulses over a defined treatment period, and cell killing is not immediate~\cite{Baskar2012}. In fact, the biological response of cells to radiation involves complex processes that unfold over time. Cells do not die instantly upon radiation exposure; instead, the effects develop gradually as DNA repair mechanisms respond to radiation-induced damage. Cells may arrest the cell cycle while trying to repair the DNA or they may induce apoptosis due to fatal DNA damage. Additionally, non-fatally damaged cells may continue to proliferate within the crowded tumour microenvironment (TME), impacting the overall effectiveness of radiotherapy~\cite{Majeed2023}. These processes are not considered in models that assume instantaneous cell killing.

To mediate this oversimplification, Wilson \emph{et al.}~\cite{Wilson2023} proposed a new death-rate approach, the CDR model. This model introduces a continuous death-rate term for radiation treatment, making the radiation effect provided by the LQ model continuous over a time period, rather than discrete and instantaneous. In the CDR model, the death rate is active for the duration of the treatment window (or treatment effectiveness window), $w$, starting at the treatment time $t_i$. Outside of this window, the death rate is zero, so that no cell killing is assumed between radiation treatments. The model creates a sequence of treatments by summing over all treatment times $t_i$ and uses a continuous approximation of the Heaviside function  to switch the treatment effect on and off while maintaining continuity of the function. The continuous death rate (CDR) for each dose fraction $d$ delivered over time window $w$ is defined as:
\begin{align}
    CDR=\sum_i V \left[ 1-e^{-\left( \alpha d+\beta d^2 \right)} \right] \frac{\hat{H} \left(t-t_i \right)-\hat{H} \left( t-\left( t_i +w \right) \right)}{w}
\end{align}
where $\hat{H}(t)$ is a continuous approximation of the Heaviside function expressed using hyperbolic tangents:
\begin{align}
    \hat{H}(t)=\frac{1}{2}\left( 1+\tanh{\frac{t}{\varepsilon}} \right)
\end{align}
with $0<\varepsilon\ll1$ to capture the steep transition of the Heaviside function. Incorporating this new death-rate term into the tumour growth equation results in the following CDR model for tumour growth with radiotherapy:
\begin{align}
    \frac{dV}{dt}=f(V)-\sum_i V \left[ 1-e^{-\left( \alpha d+\beta d^2 \right)} \right] \frac{\hat{H} \left(t-t_i \right)-\hat{H} \left( t-\left( t_i +w \right) \right)}{w}
    \label{eq:CDR}
\end{align}

\subsection{Mathematical Model for Tumour Growth}

There are many existing models of tumour growth including exponential growth, Gompertzian growth, Logistic growth, and others. For the general growth function $f(V)$ here, we use the logistic model as previous research has shown that it is effective in describing a variety of tumour types~\cite{Panetta1995}. It is a standard choice in the literature due to its balance of simplicity and physiological appropriateness. Thus, we set 
\begin{equation}
f(V)=\rho V\left( 1-\frac{V}{K} \right) \label{eq:log}
\end{equation}
where $\rho$ is the intrinsic tumour growth rate and $K$ is the carrying capacity of the tumour, representing the maximum tumour size the local environment can support. In logistic growth, the tumour volume initially grows exponentially, but the growth rate gradually slows as the volume approaches the carrying capacity.

\subsection{Experimental Data}
\begin{figure}
    \centering
    \includegraphics[width=1\linewidth]{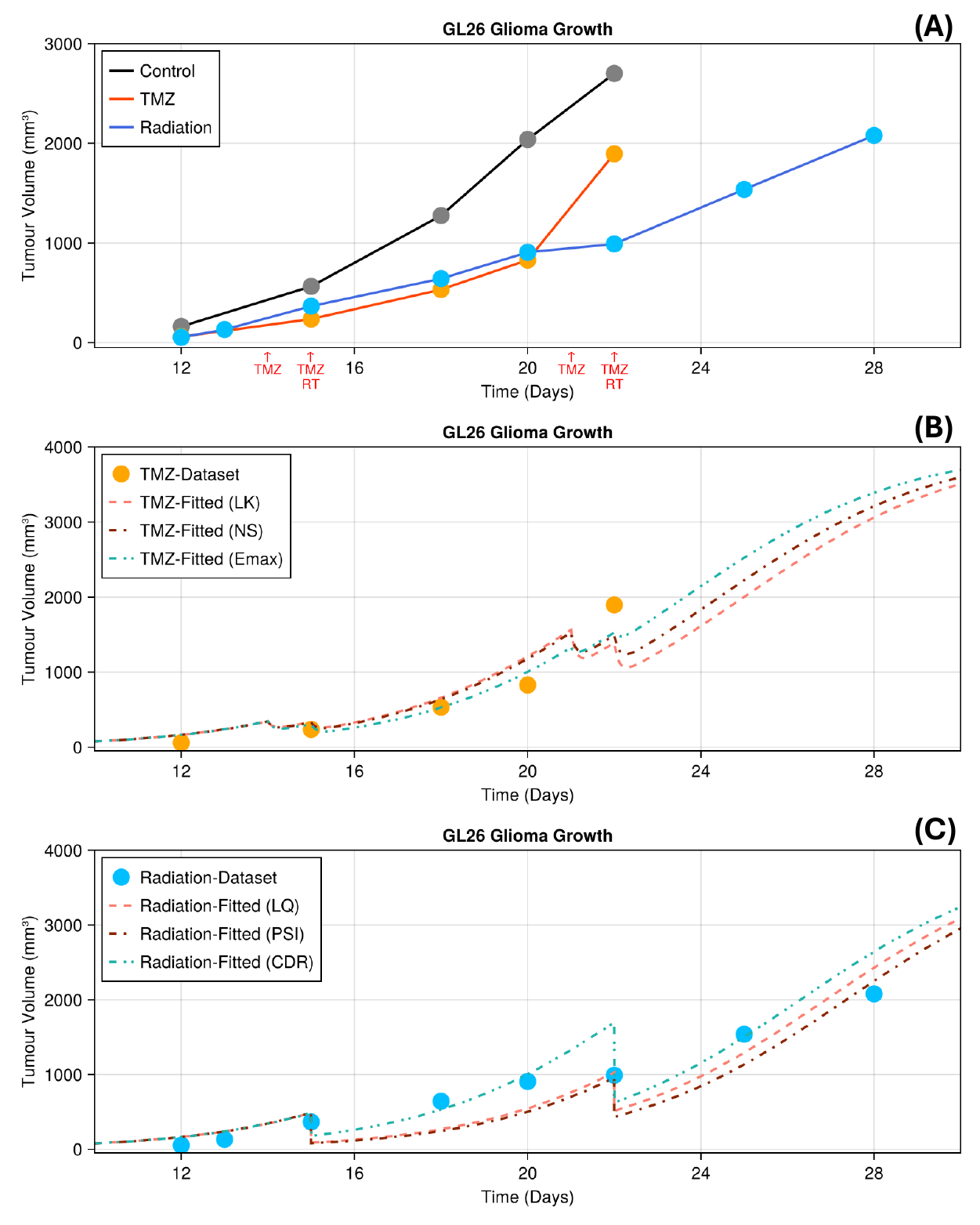}
    \caption{Subcutaneous tumour progression of murine GL26 glioma in the control group, the TMZ treatment group, and the radiation group (A),  data from~\cite{Safdie2012}. Model prediction of TMZ treatment for three chemotherapy models (B) and for three radiotherapy models (C).}
    \label{fig:dataset}
\end{figure}

To estimate model parameters, we fit each model individually to the murine glioma experimental data treated with either temozolomide chemotherapy (TMZ) or radiotherapy, presented in~\cite{Safdie2012} and shown in Figure~\ref{fig:dataset}.  The murine glioblastoma cells GL26 were injected subcutaneously with $2\times10^5$ cells into the lower back region of C57BL/6 mice. 
TMZ was delivered by injection through the lateral tail vein at $15$ mg/kg body weight. To mimic multi-cycle treatments in humans, mice were injected in two rounds for a total of $30$ mg/kg per cycle. Whole body irradiation was performed with a Cesium 137 radiation source at a dose of $5$ Gy for the first treatment and $2.5$ Gy for the second treatment. Mice were monitored daily and euthanized if displaying signs of severe stress, deteriorating health status, or a tumour burden greater than $2500$ mm\textsuperscript{3}. Tumour measurement began once the tumour became palpable under skin on day 12.

Mice $(N=12)$ in the control group were not treated with TMZ or radiation. Mice $(N=5)$ in the TMZ treatment group received $15$ mg/kg/day for two 48-hour cycles (days 13 and 15, and days 20 and 22), totaling $30$ mg/kg per cycle. Mice $(N=9)$ in the radiation treatment group were treated with $5$ Gy on day 15 and $2.5$ Gy on day 22 totaling $7.5$ Gy. The second dose was lowered to reduce radiotoxicity.

\subsection{Model Parameterization}

We use a combination of fixed and fitted model parameters using the literature when available to fix parameters, and the dataset from~\cite{Safdie2012} to estimate the remaining values with Bayesian inference. Specifically, we use Markov Chain Monte Carlo (MCMC) methods and take the Maximum Likelihood Estimate (MLE) as our best fitting parameter set for each model. 

The initial conditions for glioblastoma tumour volume in the control ($V_c(0)$) and treatment ($V_t(0)$) groups are estimated to be $2.576$ mm\textsuperscript{3} and $1.530$ mm\textsuperscript{3}, respectively, based on the dataset in~\cite{Safdie2012}. The initial condition $C(0)$ for drug concentration is set to $0$ mg/kg. Table~\ref{tab:parameterization} provides a summary of all parameter values and descriptions. 

The glioblastoma growth function is assumed to be logistic with a fixed carrying capacity of $K=4000$ mm\textsuperscript{3}. This value was chosen as it is greater than any reported measurement (all are less than $3000$ mm\textsuperscript{3}) but it is also small enough to still slow down the exponential phase of the logistic model. The growth rate $\rho$ was fit to the control dataset and estimated to be about $0.393$ day\textsuperscript{-1}. 

To estimate the drug efficacy of each chemotherapy model in~\eqref{eq:chemo}, we set the clearance rate $k$ of TMZ to be $8.318$ day\textsuperscript{-1} as reported in~\cite{Bogdanska2017}, and assume the half-maximal shape parameter $m$ of the Emax model to be $1\%$ of $K$. Each efficacy parameter $e_i$ for $i=1,2,3$ was estimated with Bayesian inference and found to be: $e_1=0.213$ kg/(mg$\cdot$day) for the LK model, $e_2=0.611$ kg/(mg$\cdot$day) for the NS model, and  $e_3=85.866$ mm\textsuperscript{3}$\cdot$kg/(mg$\cdot$day) for the Emax model. The fit of these three models to the chemotherapy treated data is shown in Figure~\ref{fig:dataset}(B).

To estimate radiosensitivity parameters for each model, we used an $\alpha / \beta$ ratio of $8$ Gy as reported in~\cite{Hingorani2012} for glioblastoma, and we fixed the CDR treatment window $w$ to 15 minutes as in~\cite{Wilson2023}. Using Bayesian inference, the remaining model parameters are estimated as  $\alpha_1=0.213$, $\beta_1=0.0266$ for the LQ model~\eqref{eq:LQ}, $\alpha_2=0.384$, $\beta_2=0.0480$ for the PSI model~\eqref{eq:PSI}, and $\alpha_3=3.784$, $\beta_3=0.469$ for the CDR model~\eqref{eq:CDR}. The fit of these three models to the radiotherapy treated data is shown in Figure~\ref{fig:dataset}(C).

\begin{table}[htbp!]
    \centering
    \caption{Summary of model parameters and their estimated values. \label{tab:parameterization}}
    \resizebox{\textwidth}{!} {
    \def\arraystretch{1.3}
    \begin{tabular}{ccccc}
    \Xhline{3\arrayrulewidth}
    \textbf{Parameter} & \textbf{Description} & \textbf{Value} & \textbf{Unit} & \textbf{Ref} \\
    \Xhline{3\arrayrulewidth}
    $\rho$ & Glioblastoma growth rate & $0.393$ & day\textsuperscript{-1} & \\
    \hline
    $K$ & Carrying capacity of glioblastoma & $4000$ & mm\textsuperscript{3} & \\
    \hline
    $e_1$ & Efficacy of TMZ (LK) & $0.213$ & kg/(mg$\cdot$day) & \\
    \hline
    $e_2$ & Efficacy of TMZ (NS) & $0.611$ & kg/(mg$\cdot$day) & \\
    \hline
    $e_3$ & Maximal Efficacy of TMZ (Emax) & $85.866$ & mm\textsuperscript{3}$\cdot$kg/(mg$\cdot$day) & \\
    \hline
    $m$ & Half-maximal shape parameter & $0.01\cdot K$ & mm\textsuperscript{3} \\
    \hline
    $k$ & TMZ clearance rate & $8.318$ & day\textsuperscript{-1} & \cite{Bogdanska2017} \\
    \hline
    $\alpha_1$ & Linear radiosensitivity parameter (LQ) & $0.213$ & Gy\textsuperscript{-1} & \\
    \hline
    $\alpha_2$ & Linear radiosensitivity parameter (PSI) & $0.384$ & Gy\textsuperscript{-1} & \\
    \hline
    $\alpha_3$ & Linear radiosensitivity parameter (CDR) & $3.784$ & Gy\textsuperscript{-1} & \\
    \hline
    $\beta_1$ & Quadratic radiosensitivity parameter (LQ) & $0.0266$ & Gy\textsuperscript{-2} & \\
    \hline
    $\beta_2$ & Quadratic radiosensitivity parameter (PSI) & $0.0480$ & Gy\textsuperscript{-2} & \\
    \hline
    $\beta_3$ & Quadratic radiosensitivity parameter (CDR) & $0.469$ & Gy\textsuperscript{-2} & \\
    \hline
    $w$ & Treatment window & $0.0104$ & day & \cite{Wilson2023} \\
    \hline
    $d$ & Treatment dose & Varied & 
    {mg/kg or Gy} & \\
    \hline
    $t_i$ & Delivery schedule of treatment & Varied & day & \\
    \hline
    $V_c(0)$ & Initial glioblastoma volume (Control) & $2.576$ & mm\textsuperscript{3} & \\
    \hline
    $V_t(0)$ & Initial glioblastoma volume (Treatment) & $1.530$ & mm\textsuperscript{3} & \\
    \hline
    $C(0)$ & Initial TMZ concentration & $0$ & mg/kg & \\
    \Xhline{3\arrayrulewidth}
    \end{tabular} }
\end{table}

\subsection{Experimental Setup}

To explore the effects of model bias in tumour growth and treatment predictions, we optimize treatment protocols for each model. For consistency, we establish the following guidelines for our computational experiments and analysis:
\begin{enumerate}[label=(\arabic*), leftmargin=*, itemsep=0pt]
    \item Numerical simulations run for 30 days with treatment constrained to start no earlier than day 5 and to finish by day 25.
    \item Potential treatment days are multiples of 5, i.e., selected from $\left\{5, 10, 15, 20, 25 \right\}$.
    \item Each treatment strategy is evaluated based on its ability to reduce the tumour volume on day 30. That is, we aim to find $\arg\min_{\text{treatment}} (V(30))$.
    \item Chemotherapy is administered at a total dose of $15\, \mathrm{mg/kg}$ delivered in three separate doses with the middle dose fixed at $5\, \mathrm{mg/kg}$.
    \item Radiotherapy is administered at a total dose of $6\, \mathrm{Gy}$ in two separate doses.
\end{enumerate}
We select day 30 as the fixed evaluation point for two clinically motivated reasons. First, treatment protocols in oncology often include predefined assessment checkpoints, such as 30-day follow-ups, to evaluate early response and inform decisions about continuing therapy, switching to another modality, or reassessing dosing strategies~\cite{Partridge2005, Fukada2018}. Although tumour regression may continue beyond this point, such intermediate evaluations are commonly used to guide clinical decision-making. Second, in some scenarios~\cite{Bi2022, Glynne2013}, the goal of treatment is not complete tumour eradication but rather sufficient volume reduction to enable safer or more effective surgical resection. In this context, assessing tumour volume at a specific time point (e.g., day 30) allows us to examine how different model assumptions influence pre-surgical treatment planning.

According to these guidelines, the possible dose delivery schedules are described in Table~\ref{tab:schedules} and the possible dose distributions are prescribed according to the following rules. Chemotherapy is delivered in three doses of size
\begin{equation}
\label{eq:chemo-lambda}
    [10(1-\lambda), 5, 10\lambda]\qquad \text{mg/kg with} \qquad \lambda \in (0, 1)
\end{equation}
over three potential treatment days. And radiation is delivered in two doses  of size
\begin{equation}
\label{eq:rad-lambda}
    [6(1-\lambda), 6\lambda] \qquad \text{Gy with} \qquad \lambda \in (0, 1)
\end{equation}
across two possible treatment days.

\begin{table}[htbp!]
    \centering
    \caption{Possible dose delivery schedules for chemotherapy and radiotherapy. \label{tab:schedules}}
    \def\arraystretch{1.3}
    \begin{tabular}{c|c|c|c|c|c|c}
    \Xhline{3\arrayrulewidth}
    \multicolumn{3}{c|}{\textbf{Chemotherapy}} & \multicolumn{4}{c}{\textbf{Radiotherapy}} \\
    \Xhline{3\arrayrulewidth}
    (5, 10, 15) & & & (5, 10) & & & \\
    (5, 10, 20) & & & (5, 15) & (10, 15) & & \\
    (5, 10, 25) & & & (5, 20) & (10, 20) & (15, 20) & \\
    (5, 15, 20) & (10, 15, 20) & & (5, 25) & (10, 25) & (15, 25) & (20, 25) \\
    (5, 15, 25) & (10, 15, 25) & & & & & \\
    (5, 20, 25) & (10, 20, 25) & (15, 20, 25) & & & & \\
    \Xhline{3\arrayrulewidth}
    \end{tabular}
\end{table}

We conduct three different numerical experiments. In the first experiment, we seek to find the optimal dose distribution, value of $\lambda$, for each model  of each treatment modality (chemotherapy or radiotherapy) across all possible dose delivery schedules. That is, for a given model of a particular treatment, we simulate all delivery schedules listed in Table~\ref{tab:schedules} for all values of $\lambda \in (0,1)$ to identify the optimal combination.  In the second experiment, using the optimal dose distribution (value of $\lambda$) identified in the first experiment, we determine the most effective protocol for sequentially applied combinations of chemotherapy and radiotherapy. In the third experiment, we determine the most effective concurrent combination of treatments - allowing chemotherapy and radiation to be delivered on the same day. 

To conclude our analysis of model bias, we conduct a global sensitivity analysis and link the results to an illustrative example of adaptive therapy. Using a clinical dataset, we examine the chemotherapy models applied to an adaptive therapy regime in terms of parameter inference and model selection.

\section{Results}
\subsection{Model Bias in Dose Distribution and Scheduling}

Using \eqref{eq:chemo-lambda} and \eqref{eq:rad-lambda} and the list of treatment schedules in Table~\ref{tab:schedules}, we explore the bias of each model individually. Specifically, we compare each model's predicted treatment efficacy on day $30$ as a function of dose distribution across the treatment days. 

\begin{sidewaysfigure}[htbp!]
    \centering
    \includegraphics[width=1\linewidth]{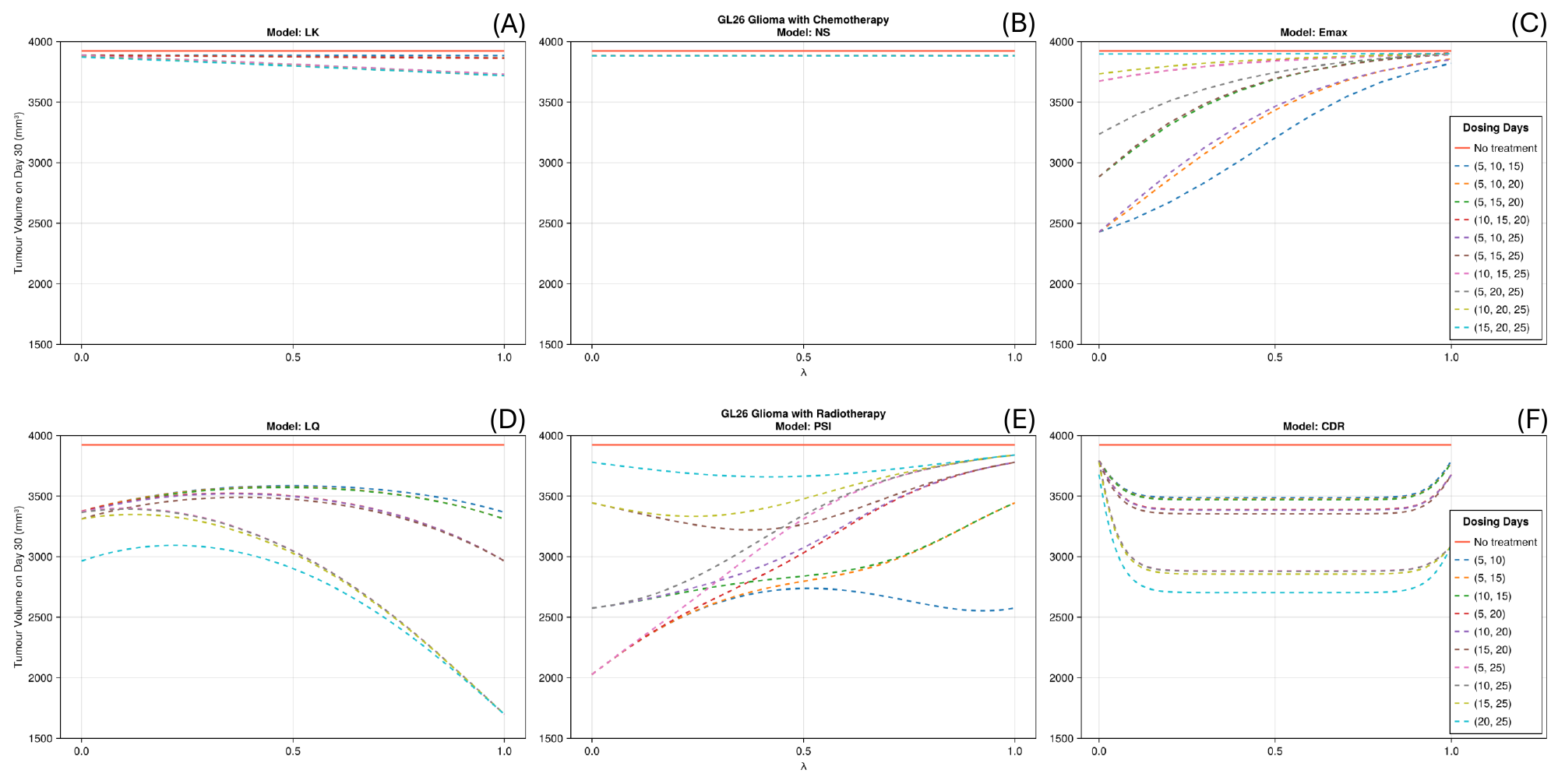}
    \caption{Tumour volume on day $30$ according to the six different models (A-F) as the dose distribution is changed ($\lambda$ on the $x$-axis), and for different treatment schedules (line colours - see legend). Recall that the dose is distributed according to $\lambda$ and corresponds to a chemotherapy schedule of $[10(1-\lambda),5,10\lambda]$ across $3$ possible days and a radiation schedule of $[6(1-\lambda,6\lambda]$ across $2$ possible days.} \label{fig:distributions}.
\end{sidewaysfigure}

Figure~\ref{fig:distributions} shows the predicted tumour volumes on day $30$ after treatment by either chemotherapy or radiation therapy for all six models discussed above. Each curve is plotted as a function of dose distribution parameter $\lambda$. As expected, all tested treatment schedules (dashed coloured lines, corresponding to Table~\ref{tab:schedules}) lead to reduced tumour volume on day $30$ compared to no treatment (solid red line), regardless of the dose distribution (value of $\lambda$). 

Chemotherapy dose is distributed across three possible treatment days according to \eqref{eq:chemo-lambda}. When $\lambda=0$, the dose distribution is $[10, 5, 0]$, concentrating doses on the first and second dosing days with no dose on the last day. This causes curves in Figure~\ref{fig:distributions} with the same first and second dosing days to start from the same point at $\lambda=0$. Similarly, when $\lambda=1$, the dose distribution is $[0, 5, 10]$, focusing doses on the second and last dosing days with no dose on the first day. Thus, curves for schedules with the same second and third dosing days converge to the same point at $\lambda=1$. 

Figure~\ref{fig:distributions} demonstrates the bias of each chemotherapy model. For the LK model, Figure~\ref{fig:distributions}A, when $\lambda=0$, the dose is front-loaded and all schedules predict a similar level of efficacy. The difference between schedules increases linearly, however, as the dose distribution shifts to back-loaded doses when $\lambda=1$. The LK model predictions show that later treatment days perform slightly better than earlier treatment days in reducing tumour volume on day $30$. Thus, the LK model predicts the smallest tumours when larger doses are delivered later, and in scheduling optimization routines, this model will bias results towards large doses delivered later on presumably larger tumour volumes.   

Surprisingly, the NS model, Figure~\ref{fig:distributions}B, displays no apparent dependence on dose distribution or schedule, suggesting it is robust to such factors. This is because the rate of cell death in the NS model is assumed to be proportional to the instantaneous growth rate, so slow tumour growth corresponds to slow drug-induced death rates, and similarly fast tumour growth corresponds to fast drug-induced death rates. In practical terms, this means a drug delivered early when the tumour is growing exponentially will have a large death rate, causing proportionally significant tumour reduction, but the early time both has a smaller overall tumour volume, and allows for regrowth before the final measurement on day $30$. Treating later, when the tumour is larger and growing slower, results in a smaller proportion of bulk reduction, but with less time to regrow. By day $30$, the two potential treatments, and in fact all simulated dose distributions and treatment schedules, have the same tumour size.

The Emax model, Figure~\ref{fig:distributions}C,  shows a large and nonlinear response to dose distribution. With front-loaded doses ($\lambda=0$), the model predicts the smallest tumours on day $30$ when the doses are delivered on the earliest possible days. The difference between predicted tumour volume for early and late treatment is quite significant. For example, compare the early schedule $(5,10,15)$ to the late schedule $(15, 20, 25)$. The difference between these curves decreases as the dose distribution shifts towards the back-end with $\lambda=1$. Thus, the Emax model performs best when large doses are delivered as early as possible. This is opposite to the bias displayed by the LK model. In scheduling optimization, the Emax model will bias results towards larger doses delivered as early as possible. 

We note also that only the NS model predicts optimal efficacy with equal dose distribution $\lambda=0.5$. Equal doses are typically administered in the clinic for simplicity of dose fractionation and to manage toxicity concerns. The LK and Emax models predict improved efficacy when the dose distribution is allowed to vary, resulting in a bias towards either back-loading or front-loading the doses, respectively. Giving larger doses, but less of them, will have toxicity management concerns that we have not considered here and leave to future work. 

Figures~\ref{fig:distributions}D-F display the predicted tumour volume on day 30 for three radiation models as a function of dose distribution, $\lambda$. Dose is distributed across two potential treatment days according to \eqref{eq:rad-lambda} and the schedules in Table~\ref{tab:schedules}. For the LQ model in Figure~\ref{fig:distributions}D, all schedules are concave down in their relation with $\lambda$, indicating that again, equalized dose fractions are not optimal at reducing tumour volume. Of course, fractionation is imposed to manage toxicity, in general, and again, we do not consider toxicity concerns at this stage. Even slight variations in the dose distribution, however, may improve efficacy -- consider say $\lambda=0.75$. The LQ model predicts the smallest tumour volumes on day $30$ when $\lambda=1$, corresponding to a dose of $6$ Gy delivered on the last treatment day only, and the schedules that have the greatest efficacy deliver this dose on the last possible treatment day (day $25$). There is a significant difference between delivering a single dose of $6$ Gy on day $20$ compared to that of day $25$ because of the tumour regrowth that occurs up to day $30$. We conclude that the LQ model has the greatest efficacy with large doses delivered as late as possible, which will bias optimization strategies towards larger and later delivery - similar to the bias demonstrated by the LK chemotherapy model. 

Figure~\ref{fig:distributions}E shows predicted tumour volume on day $30$ for the PSI model. For this model, tumour volumes vary significantly and non-linearly with dose distribution. Some tested schedules, such as $(5, 20)$ (red dashed curve), appear to be increasing with $\lambda$, indicating best efficacy is obtained when $\lambda=0$, and a dose of $6$ Gy is delivered on the first treatment day of the schedule. In contrast, other tested schedules, such as $(20, 25)$ (light blue dashed curve), are nonlinear, with a local minimum near moderate values of $\lambda$, indicating optimal efficacy with approximately equalized dose distribution. The $(5, 10)$ schedule (dark blue dashed curve) is unique in its nonlinear shape, with a local minimum occurring near $\lambda=0.9$ - a mostly back-loaded dose distribution, but its global minimum on this interval occurring at $\lambda=0$ - a front-loaded dose distribution. Because of these varying responses, it is hard to conclude on the overall bias this model may impose on optimization strategies. But we can say that the results are non-intuitive and that the model will introduce complexity into the analysis of results and model robustness, which will require extra careful interpretation. 

Conversely, the CDR model, in Figure~\ref{fig:distributions}F, seem fairly robust to dose distribution and treatment schedule. For all schedules, tumour volume decreases to a plateau centered around the equal dose distribution ($\lambda=0.5$). That is, the endpoints corresponding to front-{} or back-loaded doses perform worse than approximately equally distributed doses. The schedules that deliver doses on later days are predicted to be more effective than early treatment days, again, due to the regrowth that occurs up to day $30$.  Aligning with current practice of equalized fractionation, the CDR model may bias optimization methods towards approximately equal doses, but delivered on later treatment days.

These results depend on our choice of optimization measure. Here, we chose to optimize tumour volume on day $30$, which allows for large tumour regrowth following early treatment schedules but also large tumour growth before late treatment schedules. The growth phases may cancel each other out, or they may influence the performance shown in Figure~\ref{fig:distributions}. The choice to track tumour volume on day $30$ reflects clinical practice to implement a treatment plan and then check the results with a scan planned for some time after the round of treatment has completed.

\begin{sidewaysfigure}[htbp!]
    \centering
    \includegraphics[width=1\linewidth]{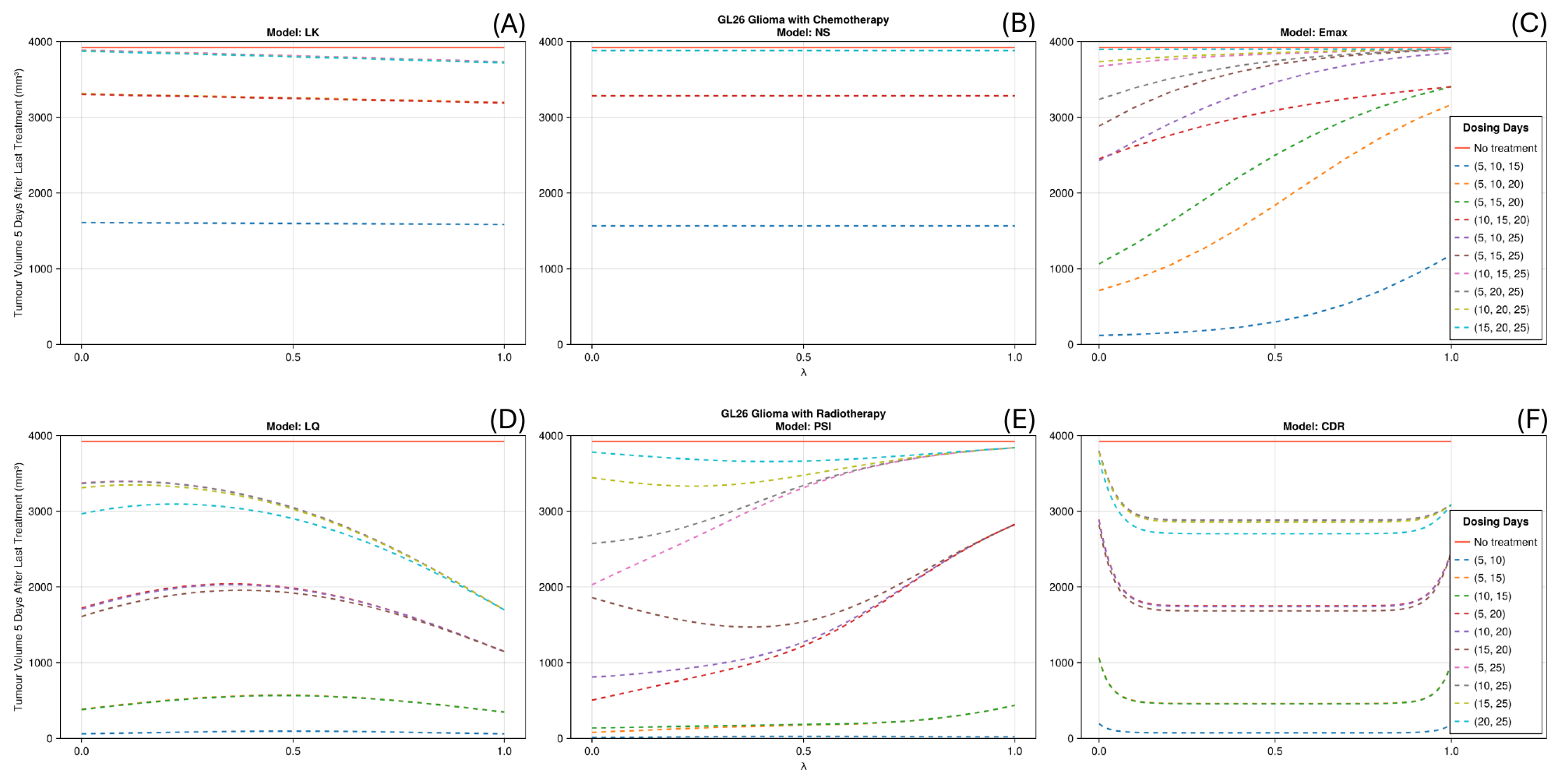}
    \caption{Tumour volume $5$ days after the last treatment day of the tested schedule according to the six different models (A-F) as the dose distribution is changed ($\lambda$ on the $x$-axis), and for different treatment schedules (line colours - see legend). The dose is distributed according to $\lambda$ and corresponds to a chemotherapy schedule of $[10(1-\lambda),5,10\lambda]$ across $3$ possible days and a radiation schedule of $[6(1-\lambda,6\lambda]$ across $2$ possible days.} 
    \label{fig:distributions2}.
\end{sidewaysfigure}

To explore this effect, the analysis was repeated but tumour volume is now tracked on the fifth day after the last treatment of the tested schedule, see Figure~\ref{fig:distributions2}. This removes the potential for large regrowth following the last treatment day. For all models, earlier treatment schedules lead to smaller tumour volumes than later treatment days. In particular, the earliest schedules consistently outperform the delayed ones, with notable differences in tumour control between early and late schedules. This demonstrates that all six models predict early treatment initiation to be most effective in suppressing tumour growth, implying that the benefit of early treatment initiation is independent of model bias and in fact reflects the underlying biology. Interestingly, among the chemotherapy models examined, the LK and NS models show little to no sensitivity to dose distribution, $\lambda$, whereas the Emax model does. Specifically, 
the Emax model becomes insensitive to dose distribution only when treatment is applied on the latest treatment days. In contrast, all radiotherapy models demonstrate sensitivity to dose distribution; however, they are all fairly robust when treatment is administered on the earliest treatment days.

\subsection{Sequential and Concurrent Protocols}

We now consider combinations of the two treatment modalities, assuming that they act independently and that their effects are additive. This is a simplification of the biology as two treatments together will typically be sub-additive in measurable effect. For sequential treatments, therapy doses are administered in a specified order on the possible schedule days taken from days 5, 10, 15, 20, and 25. We consider all possible combinations that include three chemotherapy (C) and two radiotherapy (R) treatments in series. The total number of possible sequential treatment combinations is thus 10, and these are listed in Table~\ref{tab:schedulesSeq}. 

To reduce the number of model combinations to test, we pair a model for chemotherapy with a model for radiotherapy that best matches the perceived bias found in Figure~\ref{fig:distributions}. Thus, we match the LK model with the LQ model, the Emax model with the PSI model, and the NS model with the CDR model. Each pairing consists of models that exhibit similar patterns in their response to treatment with regards to dose distribution and schedule. 

For each treatment model and each tested schedule, the optimal dose distribution for that modality is found from Figure~\ref{fig:distributions} by identifying the minimum of each curve over the closed interval $ 0.1 \le \lambda \le 0.9$ to avoid elimination of a dose. Table~\ref{tab:schedulesSeq} lists the sequential treatment schedules with the associated optimized dose distribution, value of $\lambda$, for each model, as determined for the same schedule under mono-therapy.

\begin{table}[htbp!]
    \centering
    \caption{\label{tab:schedulesSeq} Sequential multi-modal treatment protocols and their corresponding optimal dose distributions, $\lambda$ values. The optimal value of $\lambda$ for the LK, NS, and Emax chemotherapy models are denoted as $\lambda_L$, $\lambda_N$, and $\lambda_E$, respectively. And for the LQ, PSI, and CDR radiotherapy models, they are denoted as $\lambda_Q$, $\lambda_P$, and $\lambda_C$, respectively. Chemotherapy is delivered in three doses of size $[10(1-\lambda), 5, 10\lambda]$ and radiotherapy is delivered in two doses of size $[6(1-\lambda), 6\lambda]$.}
    \begin{tabular}{ccccc|cc|cc|cc}
    \toprule
    \multicolumn{5}{c}{Treatment Day} & \multicolumn{6}{c}{Model Pairs}\\
    5 & 10 & 15 & 20 & 25 & $\lambda_L$ & $\lambda_Q$ & $\lambda_E$ & $\lambda_P$ & $\lambda_N$ & $\lambda_C$ \\
    \midrule
    C & C & C & R & R & $0.9$ & $0.9$ & $0.1$ & $0.42$ & $0.5$ & $0.5$ \\
    R & C & C & C & R & $0.9$ & $0.9$ & $0.1$ & $0.1$ & $0.5$ & $0.5$ \\
    R & R & C & C & C & $0.9$ & $0.9$ & $0.1$ & $0.1$ & $0.5$ & $0.5$ \\
    C & R & R & C & C & $0.9$ & $0.9$ & $0.1$ & $0.1$ & $0.5$ & $0.5$ \\
    C & C & R & R & C & $0.9$ & $0.9$ & $0.1$ & $0.38$ & $0.5$ & $0.5$ \\
    C & C & R & C & R & $0.9$ & $0.9$ & $0.1$ & $0.24$ & $0.5$ & $0.5$ \\
    R & C & C & R & C & $0.9$ & $0.9$ & $0.1$ & $0.1$ & $0.5$ & $0.5$ \\
    C & R & C & C & R & $0.9$ & $0.9$ & $0.1$ & $0.1$ & $0.5$ & $0.5$ \\
    R & C & R & C & C & $0.9$ & $0.9$ & $0.1$ & $0.1$ & $0.5$ & $0.5$ \\
    C & R & C & R & C & $0.9$ & $0.9$ & $0.1$ & $0.1$ & $0.5$ & $0.5$ \\
  \bottomrule
    \end{tabular}
\end{table}

\begin{figure}[htbp!]
    \centering
    \includegraphics[width=1\linewidth]{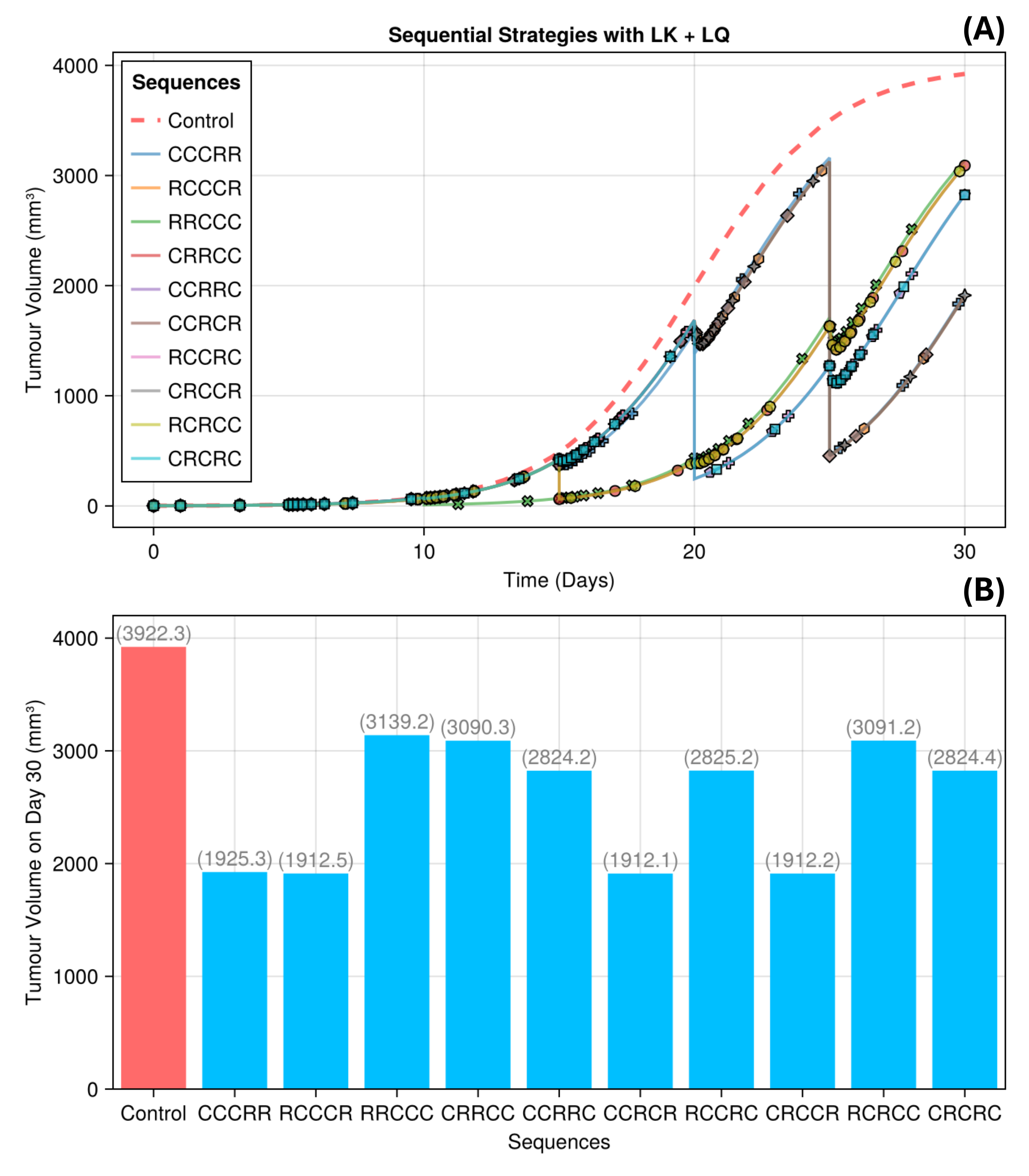}
    \caption{Tumour growth dynamics over $30$ days and the final volume on day $30$ under different sequential dosing protocols using the LK and LQ model combination.}
    \label{fig:LKLQ_seq}
\end{figure}

For each model pair, the effectiveness of the sequential protocol is assessed by measuring tumour volume on day $30$. Figure~\ref{fig:LKLQ_seq} illustrates the  growth dynamics and final tumour volume under different sequential dosing strategies for chemotherapy (C) and radiotherapy (R) using the LK + LQ model combination. In Figure~\ref{fig:LKLQ_seq}(A), the red dashed line represents tumour growth without treatment (control). All treatment sequences show varying degrees of tumour reduction compared to the control. Protocols CCCRR, RCCCR, CCRCR, and CRCCR (with  doses of $[1, 5, 9]$ mg/kg for chemotherapy and $[0.6, 5.4]$ Gy for radiation) show a sharp decline in tumour size, particularly after the final radiotherapy dose on day $25$. The LK + LQ model combination predicts that ending with radiotherapy is the most effective. This agrees with the identified bias of the LQ model, which predicts better tumour reduction for larger doses delivered later. The behaviour of the model combination is dominated by the LQ model as the radiation therapy is significantly more effective than the TMZ chemotherapy.

\begin{figure}[htbp!]
    \centering
    \includegraphics[width=1\linewidth]{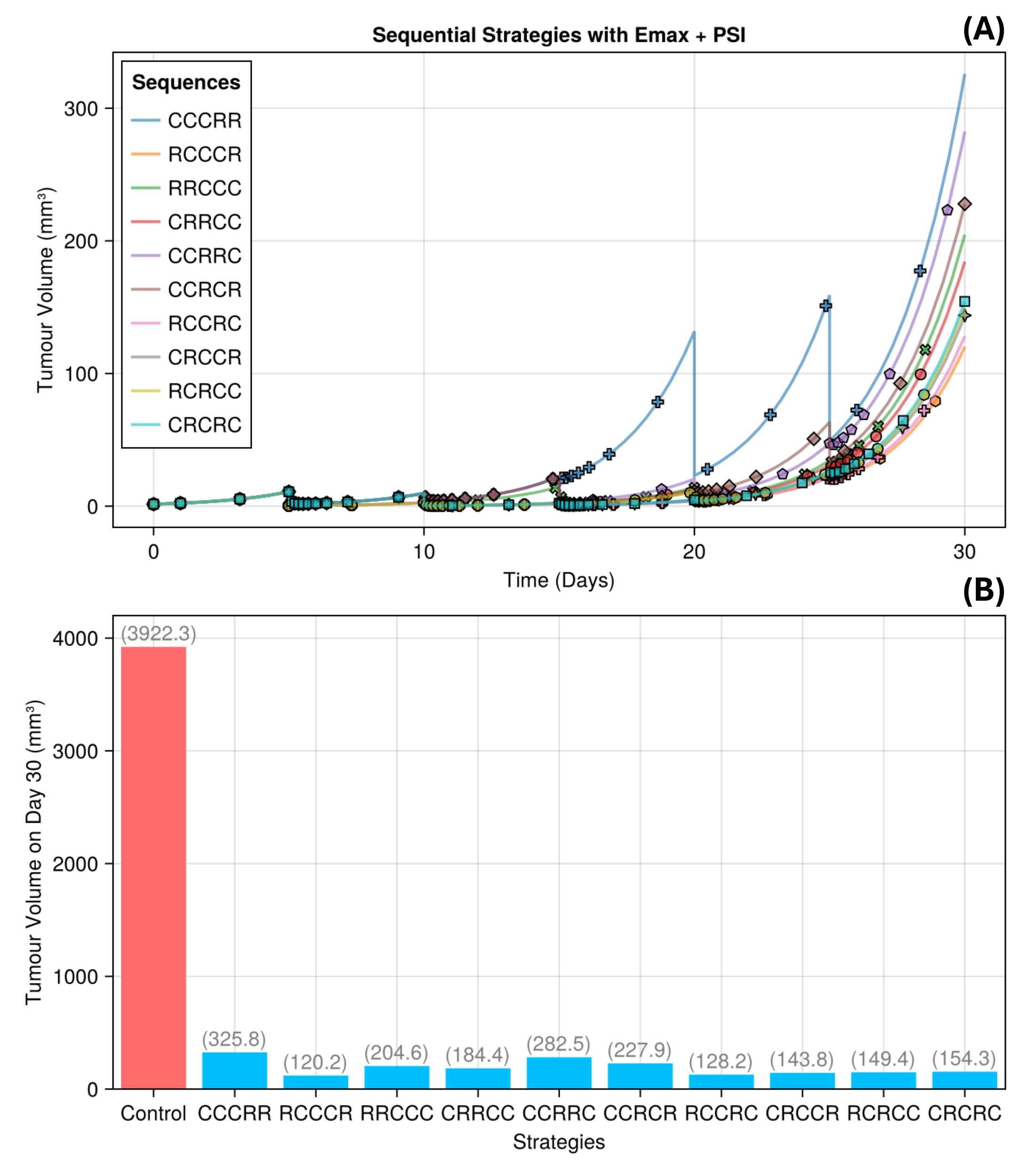}
    \caption{Tumour growth dynamics over $30$ days and the final volume on day $30$ under different sequential dosing protocols using the Emax and PSI model combination.}
    \label{fig:EmaxPSI_seq}
\end{figure}

Figure~\ref{fig:EmaxPSI_seq} shows tumour treatment dynamics for sequential dosing protocols using the Emax + PSI model combination. Compared to the LK + LQ model combination, tumour reduction is more pronounced. Sequences such as RCCCR and RCCRC, with dose distributions of $[9, 5, 1]$ mg/kg for chemotherapy and $[5.4, 0.6]$ Gy for radiotherapy, both begin with a large dose of radiation followed by two doses of chemotherapy, and result in the lowest predicted tumour volumes ($120$ mm\textsuperscript{3}). This model combination suggests that early radiotherapy effectively suppresses tumour growth and subsequent chemotherapy enhances treatment response. This is consistent with the previously identified model bias of both models, predicting smallest tumours when the largest doses are delivered as early as possible. Sequences like CCCRR and CCRRC, which start with large doses of chemotherapy, are predicted to perform the worst for this model combination, but still significantly better than the results of the previous model combination in Figure~\ref{fig:LKLQ_seq}.

\begin{figure}[htbp!]
    \centering
    \includegraphics[width=1\linewidth]{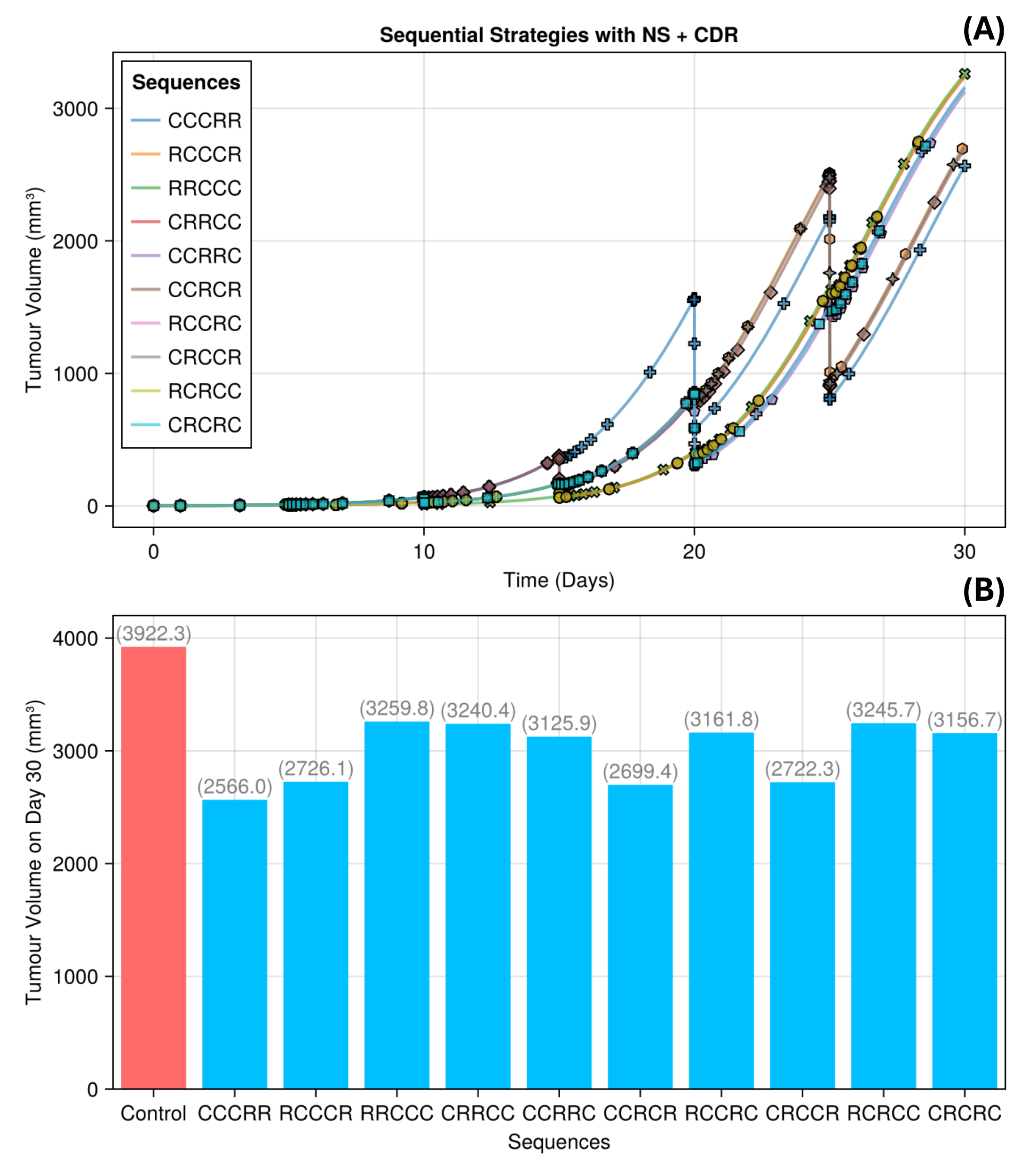}
    \caption{Tumour growth dynamics over $30$ days and the final volume on day $30$ under different sequential dosing protocols using the NS and CDR model combination.}
    \label{fig:NSCDR_seq}
\end{figure}

Finally, Figure~\ref{fig:NSCDR_seq} shows the sequential protocols using the NS + CDR model combination. For this pair of models, the optimal dose distribution is equalized: $[5, 5, 5]$ mg/kg for chemotherapy, and $[3, 3]$ Gy for radiotherapy. The models predict a slight tumour reduction advantage for protocols ending with radiotherapy, such as CCCRR and CCRCR. Specifically, the CCCRR protocol demonstrates the most pronounced tumour reduction on day $30$, slightly larger than the best protocol from the LK + LQ pairing, and much larger than all the protocols from the Emax + PSI pairing. 

For concurrent treatment protocols, we allow administration of the two treatment modalities either simultaneously or separately, on treatment days (5, 10, 15, 20, and 25). This generates $30$ possible concurrent schedules to test. Similar to the sequential analysis, we determine the optimal dose distribution for each combination using $0.1 \le \lambda \le 0.9$. Table~\ref{tab:schedulesCon} provides the list of all possible concurrent protocols along with their corresponding optimal dose distributions.

\begin{table}[tbp!]
    \centering
    \caption{\label{tab:schedulesCon} Concurrent multi-modal treatment protocols and their corresponding optimal dose distributions, $\lambda$ values. The optimal value of $\lambda$ for the LK, NS, and Emax chemotherapy models are denoted as $\lambda_L$, $\lambda_N$, and $\lambda_E$, respectively. And for the LQ, PSI, and CDR radiotherapy models, they are denoted as $\lambda_Q$, $\lambda_P$, and $\lambda_C$, respectively. Chemotherapy is delivered in three doses of size $[10(1-\lambda), 5, 10\lambda]$ mg/kg, and radiotherapy is delivered in two doses of size $[6(1-\lambda), 6\lambda]$ Gy.}
    \begin{tabular}{ccccc|cc|cc|cc}
    \toprule
    \multicolumn{5}{c}{Treatment Day} & \multicolumn{6}{c}{Model Pairs}\\
    5 & 10 & 15 & 20 & 25 & $\lambda_L$ & $\lambda_Q$ & $\lambda_E$ & $\lambda_P$ & $\lambda_N$ & $\lambda_C$ \\
    \midrule
    C & CR & CR & - & - & $0.9$ & $0.9$ & $0.1$ & $0.1$ & $0.5$ & $0.5$ \\
    CR & C & CR & - & - & $0.9$ & $0.9$ & $0.1$ & $0.1$ & $0.5$ & $0.5$ \\
    CR & CR & C & - & - & $0.9$ & $0.9$ & $0.1$ & $0.1$ & $0.5$ & $0.5$ \\
    C & CR & - & CR & - & $0.9$ & $0.9$ & $0.1$ & $0.1$ & $0.5$ & $0.5$ \\
    CR & C & - & CR & - & $0.9$ & $0.9$ & $0.1$ & $0.1$ & $0.5$ & $0.5$ \\
    CR & CR & - & C & - & $0.9$ & $0.9$ & $0.1$ & $0.1$ & $0.5$ & $0.5$ \\
    C & CR & - & - & CR & $0.9$ & $0.9$ & $0.1$ & $0.1$ & $0.5$ & $0.5$ \\
    CR & C & - & - & CR & $0.9$ & $0.9$ & $0.1$ & $0.1$ & $0.5$ & $0.5$ \\
    CR & CR & - & - & C & $0.9$ & $0.9$ & $0.1$ & $0.1$ & $0.5$ & $0.5$ \\
    C & - & CR & CR & - & $0.9$ & $0.9$ & $0.1$ & $0.38$ & $0.5$ & $0.5$ \\
    CR & - & C & CR & - & $0.9$ & $0.9$ & $0.1$ & $0.1$ & $0.5$ & $0.5$ \\
    CR & - & CR & C & - & $0.9$ & $0.9$ & $0.1$ & $0.1$ & $0.5$ & $0.5$ \\
    C & - & CR & - & CR & $0.9$ & $0.9$ & $0.1$ & $0.24$ & $0.5$ & $0.5$ \\
    CR & - & C & - & CR & $0.9$ & $0.9$ & $0.1$ & $0.1$ & $0.5$ & $0.5$ \\
    CR & - & CR & - & C & $0.9$ & $0.9$ & $0.1$ & $0.1$ & $0.5$ & $0.5$ \\
    C & - & - & CR & CR & $0.9$ & $0.9$ & $0.1$ & $0.42$ & $0.5$ & $0.5$ \\
    CR & - & - & C & CR & $0.9$ & $0.9$ & $0.1$ & $0.1$ & $0.5$ & $0.5$ \\
    CR & - & - & CR & C & $0.9$ & $0.9$ & $0.1$ & $0.1$ & $0.5$ & $0.5$ \\
    - & C & CR & CR & - & $0.9$ & $0.9$ & $0.1$ & $0.38$ & $0.5$ & $0.5$ \\
    - & CR & C & CR & - & $0.9$ & $0.9$ & $0.1$ & $0.1$ & $0.5$ & $0.5$ \\
    - & CR & CR & C & - & $0.9$ & $0.9$ & $0.1$ & $0.1$ & $0.5$ & $0.5$ \\
    - & C & CR & - & CR & $0.9$ & $0.9$ & $0.1$ & $0.24$ & $0.5$ & $0.5$ \\
    - & CR & C & - & CR & $0.9$ & $0.9$ & $0.1$ & $0.1$ & $0.5$ & $0.5$ \\
    - & CR & CR & - & C & $0.9$ & $0.9$ & $0.1$ & $0.1$ & $0.5$ & $0.5$ \\
    - & C & - & CR & CR & $0.9$ & $0.9$ & $0.1$ & $0.42$ & $0.5$ & $0.5$ \\
    - & CR & - & C & CR & $0.9$ & $0.9$ & $0.1$ & $0.1$ & $0.5$ & $0.5$ \\
    - & CR & - & CR & C & $0.9$ & $0.9$ & $0.1$ & $0.1$ & $0.5$ & $0.5$ \\
    - & - & C & CR & CR & $0.9$ & $0.9$ & $0.1$ & $0.42$ & $0.5$ & $0.5$ \\
    - & - & CR & C & CR & $0.9$ & $0.9$ & $0.1$ & $0.24$ & $0.5$ & $0.5$ \\
    - & - & CR & CR & C & $0.9$ & $0.9$ & $0.1$ & $0.38$ & $0.5$ & $0.5$ \\
    \bottomrule
    \end{tabular}
\end{table}

\begin{sidewaysfigure}[htbp!]
    \centering
    \includegraphics[width=1\linewidth]{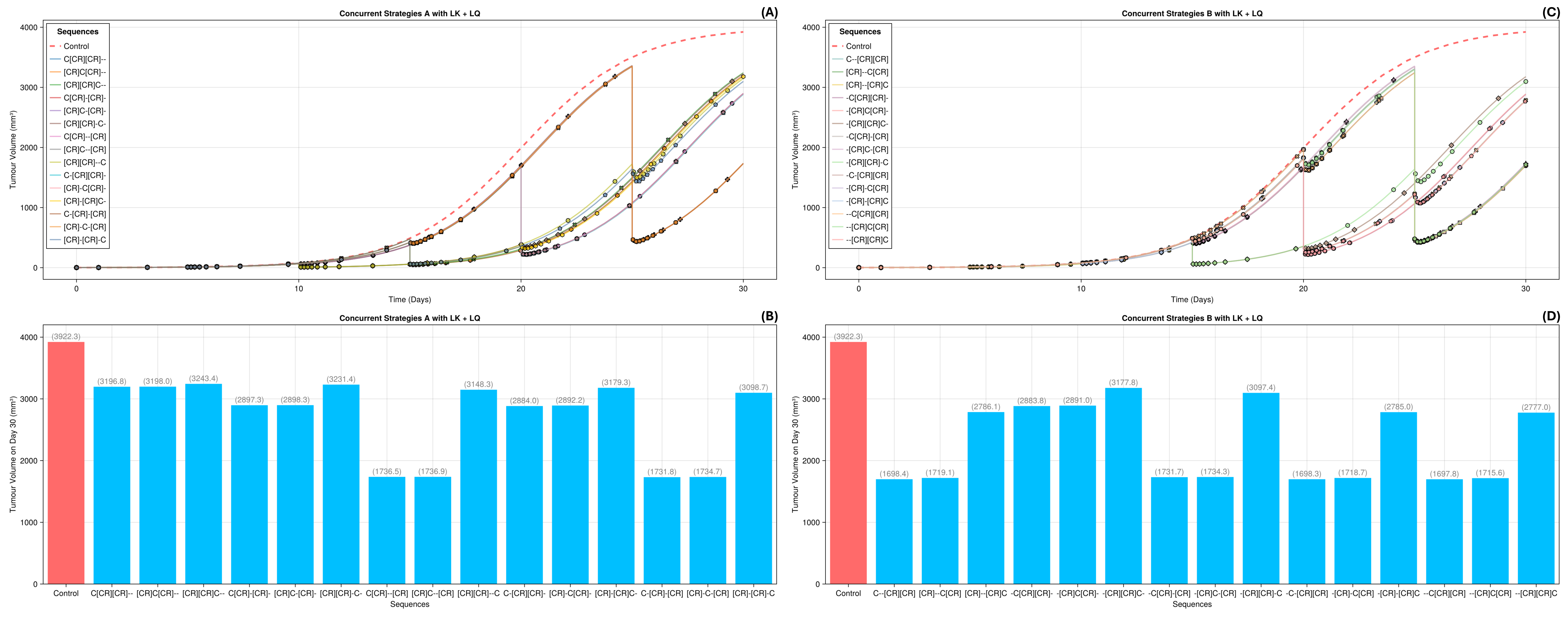}
    \caption{Tumour growth dynamics over $30$ days and the final volume on day $30$ under different concurrent dosing protocols using the LK and LQ model combination.}
    \label{fig:LKLQ_con}
\end{sidewaysfigure}

Figure~\ref{fig:LKLQ_con} shows the predicted tumour dynamics over $30$ days (top row) and final volumes on day $30$ (bottom row) under different concurrent dosing protocols for chemotherapy and radiotherapy using the LK + LQ model combination. For all tested protocols, doses are distributed according to $[1, 5, 9]$ mg/kg for chemotherapy and $[0.6, 5.4]$ Gy for radiation. Protocols that conclude with a large concurrent dose of chemotherapy and radiation, such as C--[CR][CR], predict the most tumour reduction, consistent with the biased preference of these models for late delivery of large treatment doses. In contrast, sequences such as [CR][CR]C--, which administer the combination dose early show limited effectiveness by day $30$. 

\begin{sidewaysfigure}[htbp!]
    \centering
    \includegraphics[width=1\linewidth]{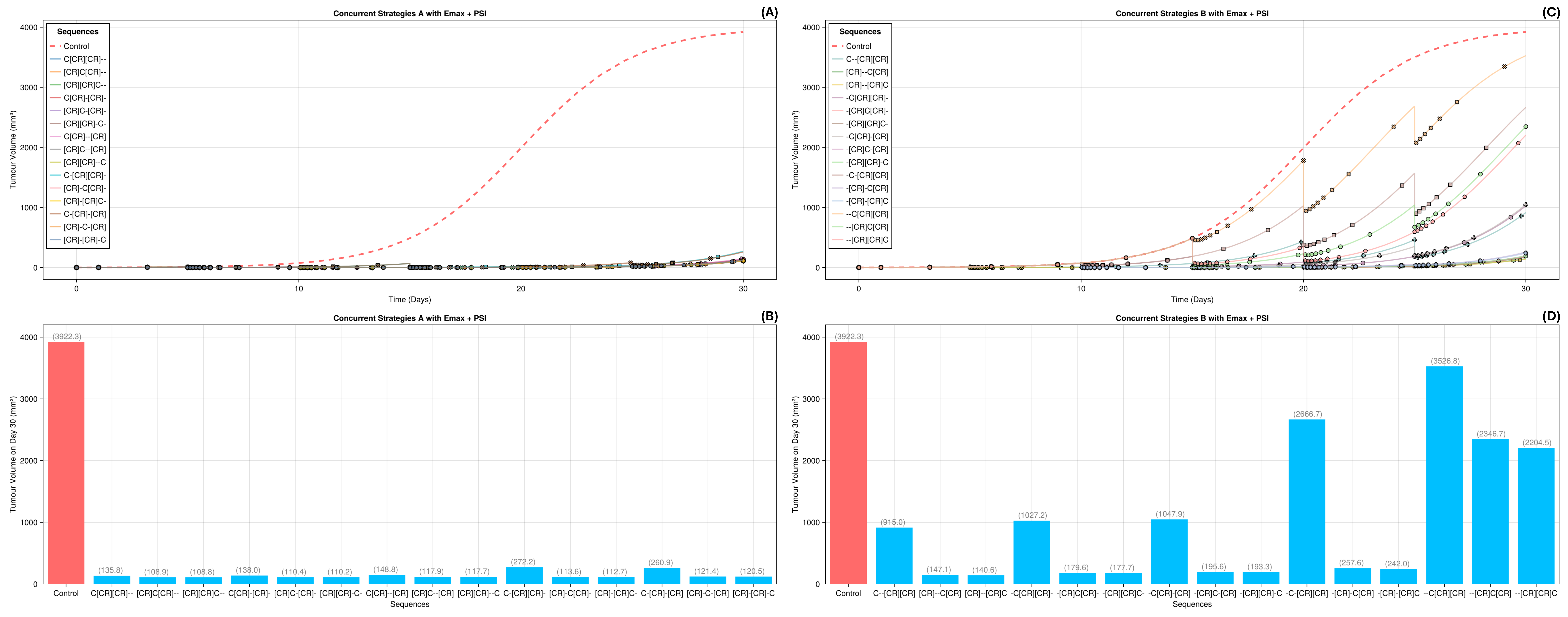}
    \caption{Tumour growth dynamics over $30$ days and the final volume on day $30$ under different concurrent dosing protocols using the Emax and PSI model combination.}
    \label{fig:EmaxPSI_con}
\end{sidewaysfigure}

Figure~\ref{fig:EmaxPSI_con} shows the results of the concurrent protocols for the Emax and PSI model combination. Optimized dose distributions for chemotherapy are $[9, 5, 1]$ mg/kg and are mostly $[5.4, 0.6]$ Gy for radiation, although some protocols deviate from this distribution as in Table~\ref{tab:schedulesCon}. Sequences that begin with a concurrent dose of chemo-radiation on either the first or second possible treatment day show the largest tumour reduction. This agrees with the bias seen for these models individually and in sequential dosing: for improved efficacy treat with large doses delivered as early as possible. Sequences ending with two consecutive chemo-radiation doses are the least effective. Again, this model combination predicts significant tumour reduction compared to the LK + LQ model in Figure~\ref{fig:LKLQ_con}.

\begin{sidewaysfigure}[htbp!]
    \centering
    \includegraphics[width=1\linewidth]{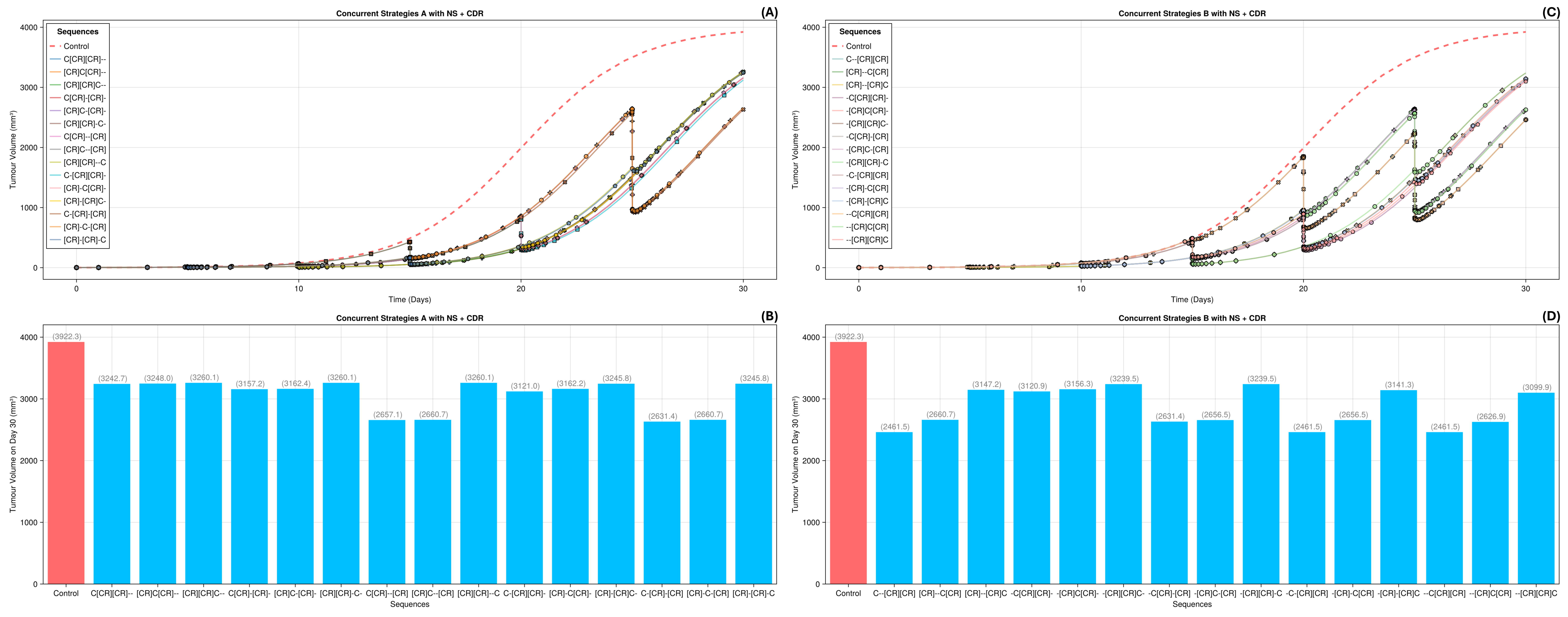}
    \caption{Tumour growth dynamics over $30$ days and the final volume on day $30$ under different concurrent dosing protocols using the NS and CDR model combination.}
    \label{fig:NSCDR_con}
\end{sidewaysfigure}

Figures~\ref{fig:NSCDR_con} shows the tumour treatment dynamics over $30$ days (top row) and the final volume on day $30$ (bottom row) for all concurrent dosing protocols and the NS + CDR model combination. Dose is distributed equally for these models, with chemotherapy doses of $[5, 5, 5]$ mg/kg and radiation doses of $[3, 3]$ Gy. Protocols that end with chemo-radiation combinations on the last possible day predict the smallest final volumes. This agrees with the model bias of the radiation CDR model, which is the stronger of the two treatment effects.

\begin{figure}[htbp!]
    \centering
    \includegraphics[width=1\linewidth]{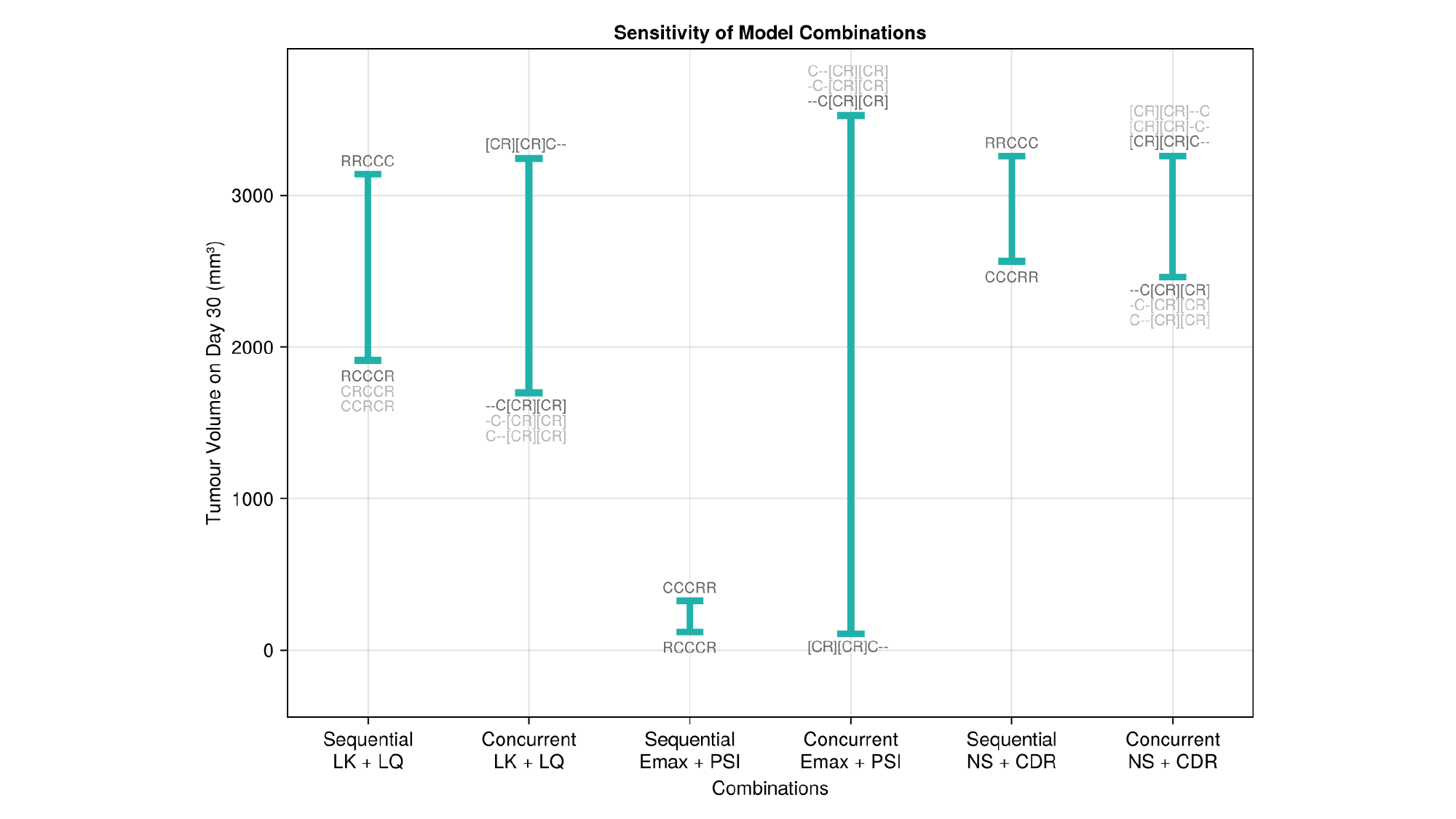}
    \caption{The range of predicted tumour volumes on day 30 for various model combinations (LK + LQ, Emax + PSI, and NS + CDR) for sequential and concurrent dosing of chemotherapy and radiotherapy.}
    \label{fig:sensitivity}
\end{figure}

An overall summary of predicted final tumour volumes on day $30$ is shown in Figure~\ref{fig:sensitivity} for each pair of models with either sequential or concurrent treatment. The best (bottom of range) and worst (top of range) treatment protocols are indicated in grey. The Emax + PSI model pair can predict significantly better tumour reduction than the other model pairs. However, while this prediction is quite robust for sequential treatments, the model pair is extremely sensitive to the applied schedule for concurrent protocols. This is demonstrated by the very narrow range for sequential, and very wide range for concurrent, protocols under the Emax + PSI model combination in Figure~\ref{fig:sensitivity}. The NS + CDR model combination both optimize to equalized dose distributions, which has the advantage that it matches typical clinical practice. The model pair is also robust to treatment schedule for both sequential and concurrent protocols. Notably, the optimal concurrent schedules for the LK + LQ and NS + CDR model combinations (two chemo-radiation treatments of increasing dose, or of equal dose, respectively delivered as late as possible) contradict the optimal schedule for the Emax + PSI model pair (two chemo-radiation treatments of decreasing dose delivered as early as possible). These findings help to demonstrate that model functional forms significantly influence the optimized treatment schedule. And that because of this model bias, different models fit to the same experimental data, can, in fact, predict contradictory optimal treatment protocols. 

\subsection{Adaptive Therapy to Demonstrate Impact of Model Bias}

Adaptive therapy (AT) is a novel approach to cancer treatment that adjusts drug administration dynamically based on the tumour's response to therapy~\cite{Hansen2020}. Unlike traditional protocols that rely on pre-determined schedules and maximum tolerated doses targeted at eradicating the tumour, AT is a non-curative approach that attempts to maintain an approximately stable tumour burden through competition between therapy-sensitive and therapy-resistant cancer cells~\cite{Gatenby2009}. By introducing treatment pauses, AT allows a portion of sensitive cells to survive, effectively suppressing resistant cells through competition. This evolutionary-based strategy tries to prolong the effectiveness of treatment and delay or prevent treatment failure due to the emergence of resistance~\cite{Hansen2020}. 


We use a simplified simulation of AT that aligns with our previous analysis and experimental data, to demonstrate the consequences of bias in treatment models. We note that AT is not appropriate for glioma treatment, but use our parameter estimates previously found for demonstration purposes only. To simplify the AT protocol, we first only consider treatment-sensitive cells as defined by equations \eqref{eq:growth} and \eqref{eq:drug}, to clearly illustrate the effect of the chosen models. First, we administer a series of three chemotherapy doses and two radiation doses, using both equal dosing and the optimized dosing from Table~\ref{tab:schedulesSeq}, combined with the optimal concurrent schedule determined in the previous section. Simplifying the idea of range-bounded adaptive therapy in~\cite{Brady2022}, our multi-dose protocol is applied whenever the tumour volume exceeds a threshold of $1000$ mm\textsuperscript{3}. Crossing this threshold triggers delivery of three doses according to the optimal concurrent schedule for the model combination. This protocol is also compared to an equal dose ($\lambda=0.5$) distribution on the three treatment days. All treatments are stopped by day 50, and the final tumour volume is measured on day 55.

\begin{figure}[htbp!]
    \centering
    \includegraphics[width=0.8\textwidth]{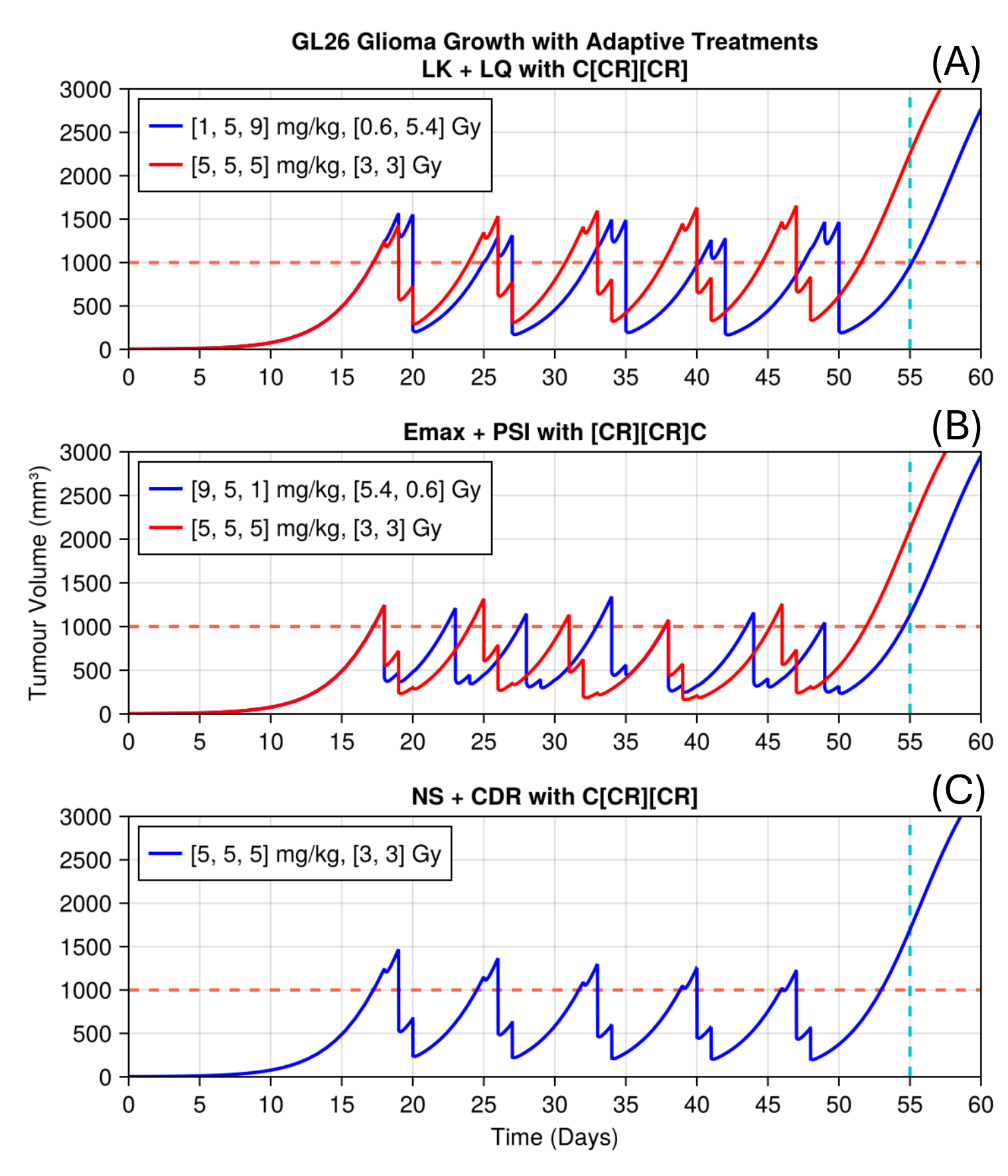}
    \caption{Simplified adaptive therapy simulations with optimized and equal dose distributions for our three model combinations. Only sensitive cancer cells are considered according to equations \eqref{eq:growth} and \eqref{eq:drug}.}
    \label{fig:adaptive}
\end{figure}

Figure~\ref{fig:adaptive} displays the results of our simplified AT. Differences appear between the triggered treatments depending on the choice of model combinations and the dose distribution. Specifically, under optimized dosing (blue curves), the LK + LQ (with increasing doses) and NS + CDR (with equal doses) predict five rounds of treatment, while the Emax + PSI (with decreasing doses) predict seven rounds. In contrast, when equal dose distributions ($\lambda = 0.5$) are applied, all model combinations predict five rounds of treatment.

We now consider the existence of both treatment-sensitive and resistant tumour cells with our framework for simplified AT as above. With sensitive ($S$) and resistant ($R$) cells, a simple model for their combined growth with competition is~\cite{Gevertz2025}:
\begin{align}
    \frac{dS}{dt}&=\rho_S S\left(1-\frac{S+R}{K}\right) \label{eq:sensitive} \\
    \frac{dR}{dt}&=\rho_R R\left(1-\frac{\beta S+R}{K}\right) \label{eq:resistant}
\end{align}
where $\rho_S$ and $\rho_R$ are the intrinsic growth rates of the sensitive and resistant tumour cells, respectively (here, we assume $\rho_S = \rho_R$ for simplicity). The parameter $\beta$ is the competition coefficient allowing $S$ to suppress the growth of $R$, and $K$ is the carrying capacity of the total tumour. For these parameters, we set $\rho_S = \rho_R = 0.393\,\mathrm{day^{-1}}$, $\beta = 5$, and $K = 4000\,\mathrm{mm^3}$. As in the first example, we administer a series of three chemotherapy and two radiotherapy treatments once the total tumour volume ($S+R$) exceeds a threshold of $1000$ mm\textsuperscript{3}, using both equal dosing and optimal dosing for each model combination. These are combined with the optimal concurrent schedule, and treatment is stopped once the total tumour volume reaches $3000$ mm\textsuperscript{3}. The day on which this volume is reached is termed the time of treatment failure, as the resistant cells lead to tumour progression.

\begin{figure}[htbp!]
    \centering
    \includegraphics[width=0.8\textwidth]{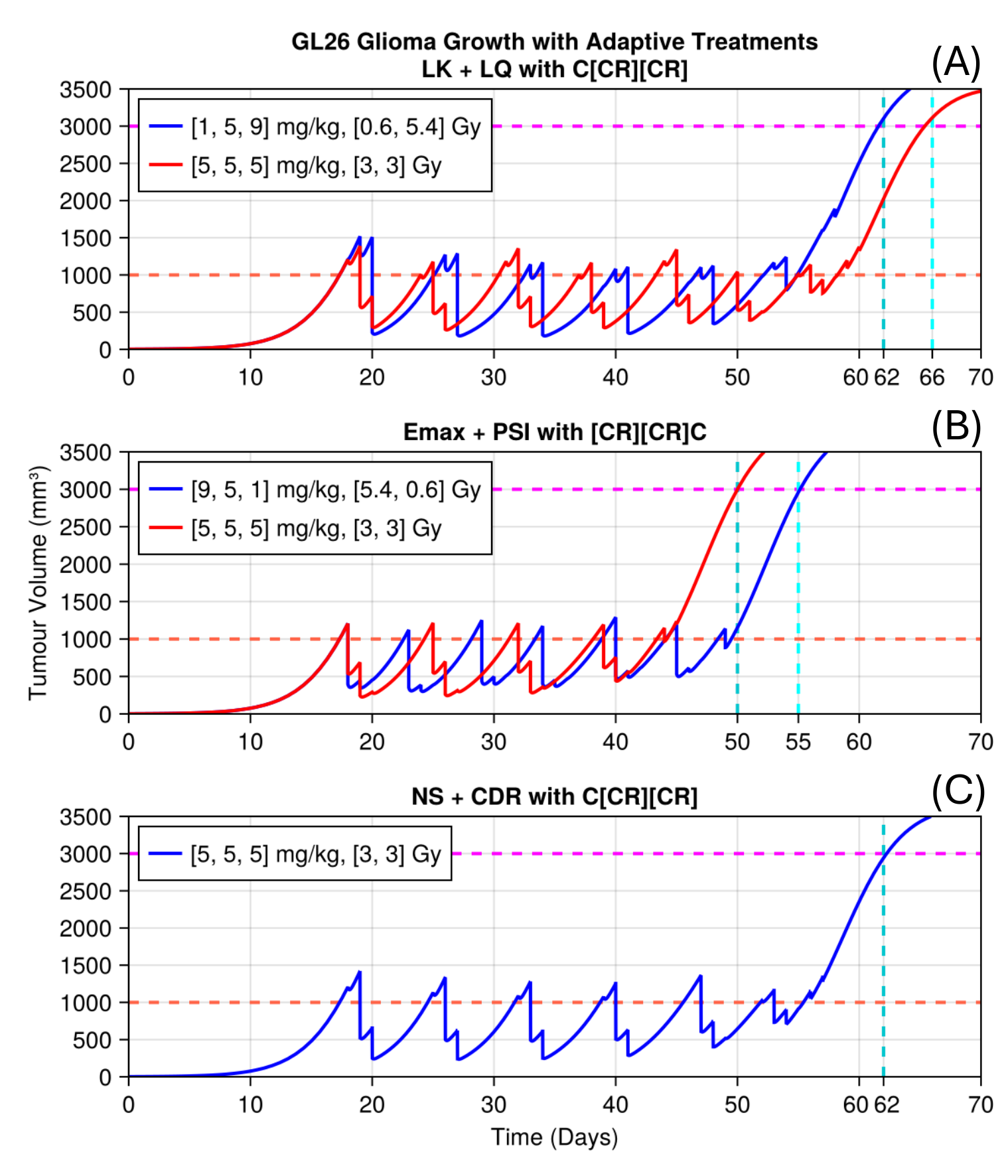}
    \caption{Adaptive treatment simulations with optimal treatment strategies for therapy-sensitive and therapy-resistant tumour cells according to equations \eqref{eq:sensitive} and \eqref{eq:resistant}.}
    \label{fig:adaptive2}
\end{figure}

Figure~\ref{fig:adaptive2} displays the results of our second AT example. Due to the presence of resistant cells, the adaptive treatment ultimately fails. In the LK + LQ and NS + CDR model combinations, the first five cycles effectively reduce tumour volume, but the efficacy of the sixth cycle is diminished. On the other hand, in the Emax + PSI model combination with optimized dosing (decreasing doses), the first six cycles successfully reduce tumour burden below the treatment-triggering threshold, but the seventh does not. Under equal dosing, the first four cycles suppress tumour regrowth, but failure occurs during the fifth cycle. Furthermore, the Emax + PSI model 
predicts treatment failure around days $50-55$ while the other two model pairs predict failure around days $62-66$. These differences in number of treatment cycles, overall delivered dose, and time to failure may be exaggerated, however, for different parameter value combinations and longer periods of treatment cycling. 

These AT examples demonstrate how the model predictions are influenced by the inherent bias within each model, and how personalized medicine, and, as an example, AT, are highly sensitive to model selection. Overall, this highlights the need for careful analysis of inherent model biases and assumptions, full calibration and validation of the models, and extensive uncertainty quantification of the predicted results, especially for personalized therapy and digital twins.

\subsection{Global Sensitivity Analysis of Treatment Models}

To examine how parameter variability influences treatment predictions, we conducted a global sensitivity analysis using the Sobol method on all models. For chemotherapy models (LK, NS, Emax), treatment consisted of three equal doses of 5 mg/kg administered on days 10, 15, and 20. Tumour volume was evaluated on day 30. The parameters varied were tumour growth rate ($\rho$), carrying capacity ($K$), drug clearance rate ($k$), and drug efficacy ($e$). Ranges were defined relative to the fitted values in Table 1, with $\rho$, $K$, and $k$ varied from one-half to twice their baseline values, and $e$ varied from one-twentieth to twenty times the baseline value, to capture weak sensitivity.

For radiotherapy models (LQ, PSI, CDR), treatment consisted of two equal fractions of 3 Gy delivered on days 10 and 20 with tumour volume assessed on day 30. The parameters varied were $\rho$, $K$, and the radiosensitivity parameter $\alpha$. As in chemotherapy, $\rho$ and $K$ were varied from one-half to twice their baseline values, and $\alpha$ was varied from one-twentieth to twenty times the baseline.

\begin{figure}[htbp!]
    \centering
    \includegraphics[width=1\linewidth]{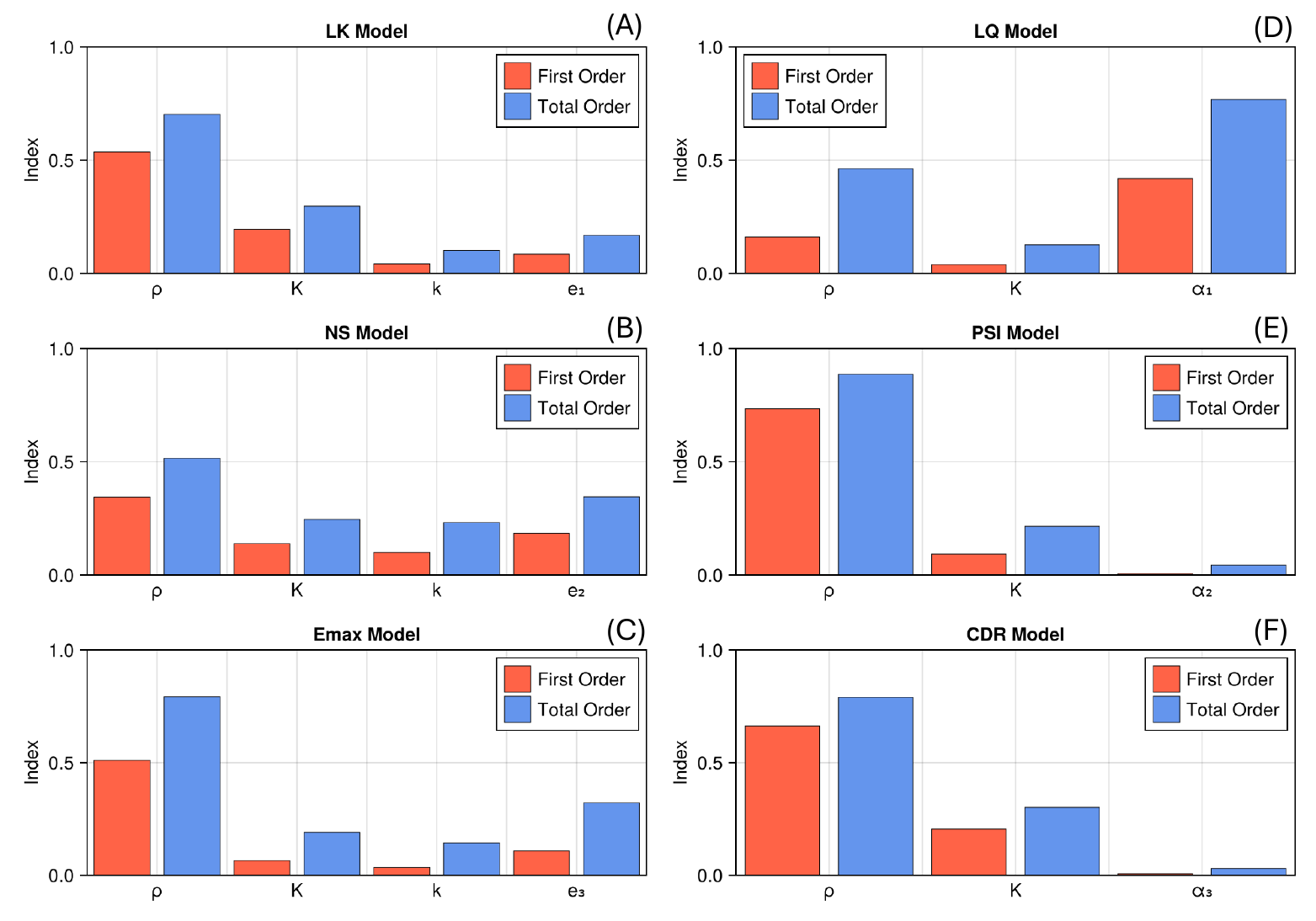}
    \caption{Global sensitivity analysis of chemotherapy models (A) LK, (B) NS, and (C) Emax, and radiotherapy models (D) LQ, (E) PSI, and (F) CDR, using the Sobol method. Bars represent first-order (red) and total-order (blue) Sobol indices for key model parameters. Parameters assessed for the chemotherapy models are the tumour growth rate, $\rho$, the carrying capacity, $K$, the drug clearance rate, $k$, and the drug efficacy, $e$. Parameters assessed for the radiotherapy models are $\rho$, $K$, and the radiosensitivity parameter $\alpha$.}
    \label{fig:sensitivity_sobol}
\end{figure}

The sensitivity results are shown in Figure~\ref{fig:sensitivity_sobol}. For chemotherapy models (A–C), growth rate, $\rho$, is the most influential parameter, dominating both first-order and total-order indices. Carrying capacity, $K$, contributes moderately in the LK and NS models. In contrast, drug clearance rate, $k$, shows minimal influence in all models. By total-order, drug efficacy, $e$, ranks second in significance in the NS and Emax models, and third in the LK model. 

For radiotherapy models (D–F), radiosensitivity, $\alpha$, is ranked most sensitive in the LQ model but least sensitive in the PSI and CDR models, where the radiation killing effects are moderated by the growth function or the treatment time, respectively. In these two models, growth rate, $\rho$, is the most significant parameter. The reduced significance of $\alpha$ suggests that the PSI and CDR models are more robust to variation in radiosensitivity estimates than the LQ model.



\subsection{Clinical Parameterization of Chemotherapy Models}

To assess model suitability to real-world clinical data, we used the prostate cancer dataset from~\cite{Bruchovsky2007}, publicly available at \url{http://www.nicholasbruchovsky.com/clinicalResearch.html}. This prostate cancer clinical dataset was recently analyzed in~\cite{Strobl2022}. Each patient file contains longitudinal observations with ten variables: (1) patient number, (2) observation date, (3) Cyclophosphamide (CPA) dose, (4) Leuprolide (LEU) dose, (5) prostate-specific antigen (PSA) level, (6) testosterone level, (7) cycle number, (8) treatment status, (9) day number, and (10) an alternative day number.

We seek to determine the distribution of drug efficacy parameter values for the patient population. We considered only three of the reported variables: CPA dose, PSA level, and the alternative day number. PSA is a serum biomarker routinely used in prostate cancer management and was taken here as a proxy for tumour burden in our model. This assumption follows the earlier modelling study~\cite{Strobl2022}, where PSA dynamics are used as a surrogate for measuring growth and treatment dynamics. CPA is an alkylating agent widely used in prostate cancer therapy.

In the dataset, patients were excluded if CPA records were missing or inconsistent with observed PSA decline. Only CPA was considered as the treatment variable because its administration was strongly associated with reductions in PSA. After these exclusions, 65 patients remained for analysis, consistent with the prior study~\cite{Strobl2022}. To allow comparison across patients, PSA values were normalized to one, relative to the baseline measurement at the start of therapy.

The sensitive/resistant subpopulation tumour model, equations \eqref{eq:sensitive} and \eqref{eq:resistant}, was applied to the dataset with simulated chemotherapy affecting only the sensitive cells. CPA efficacy ($e$) was inferred using MCMC both at the individual level (each patient is fit separately) and at the cohort level (all patients are fit simultaneously) for each chemotherapy model (LK, NS, Emax). The remaining parameter values were fixed: $\rho_S = 0.027\,\mathrm{day^{-1}}$, $\rho_R = 0.00594\,\mathrm{day^{-1}}$, and $\beta = 1$ from~\cite{Strobl2022}, $k = 0.308\,\mathrm{day^{-1}}$ from~\cite{Kuhnz1993}, and $K = 30000\,\mathrm{mm^3}$.

To help inform model selection, that is, which chemotherapy model is most supported by the clinical data at the individual level, the Bayesian information criterion (BIC) was computed for each patient using the LK, NS, and Emax models. The BIC is defined as
\begin{align}
    \text{BIC}=k\ln{n}-2\ln{\hat{L}}
\end{align}
where $k$ is the number of estimated parameters, $n$ is the number of data points, and $\hat{L}$ is the maximum likelihood. Lower BIC values indicate a better balance between goodness of fit and model simplicity.

\begin{figure}[htbp!]
    \centering
    \includegraphics[width=0.9\linewidth]{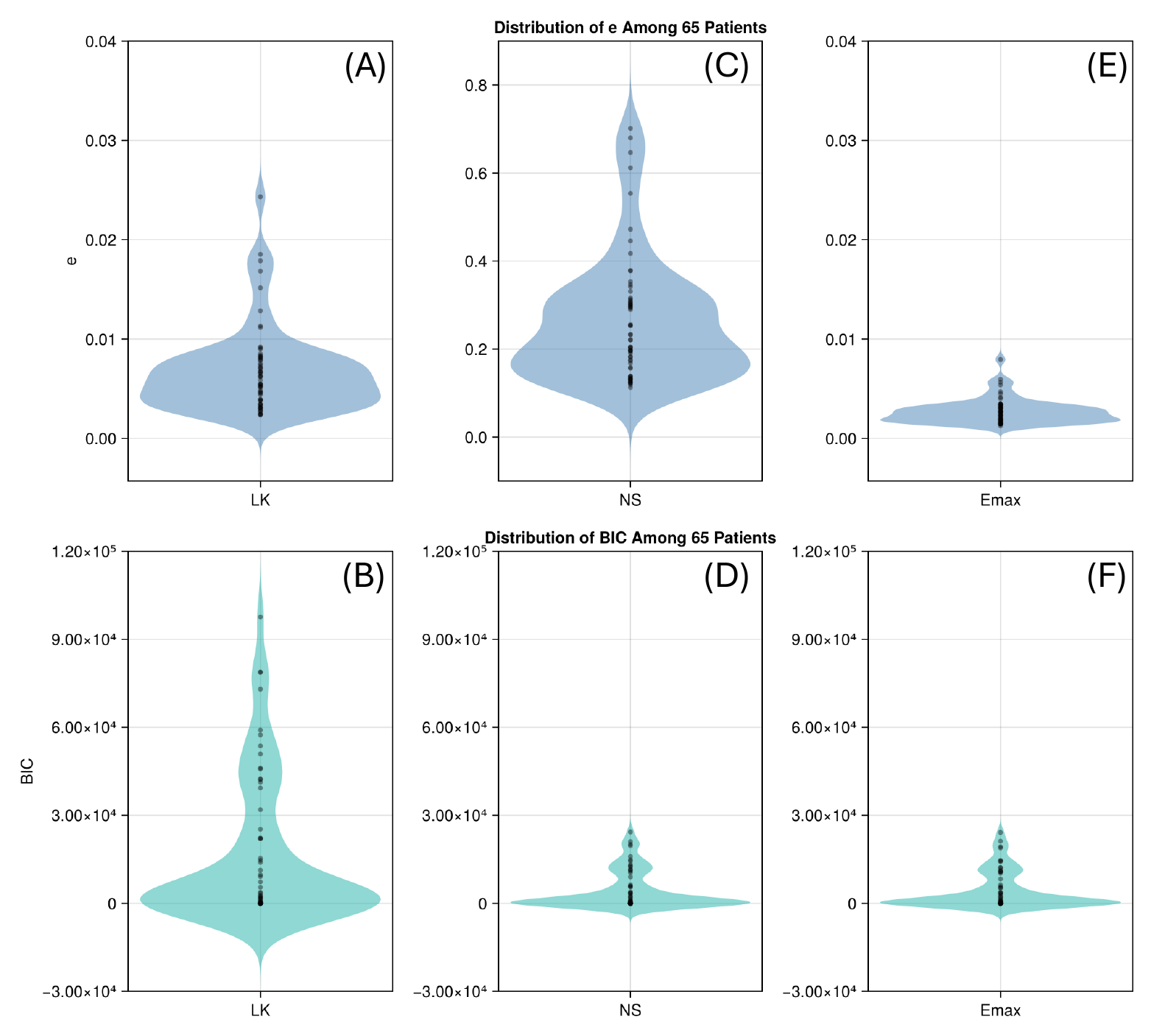}
    \caption{Distributions of patient-specific CPA efficacy parameter values $(e)$ inferred with MCMC and the corresponding BIC values for each of the 65 patients and for each of the three chemotherapy models: LK (A, B), NS (C, D), and Emax (E, F). The upper panels show violin plots of the aggregated patient-specific estimates of $e$, and the lower panels show the corresponding distributions of BIC values.}
    \label{fig:violinplot}
\end{figure}

Figure~\ref{fig:violinplot} shows the result of the individual-based parameter inference for CPA efficacy and BIC. The violin plots for CPA efficacy using the (A) LK, (C) NS, and (E) Emax models all have a “spinning-top” appearance, but across different ranges of the parameter value. The distribution of inferred efficacy values for the Emax model is the most compact, LK is slightly wider, and NS is much wider with a heavier tail. This suggests that the Emax model is able to represent the variability in the dataset with the smallest range of parameter values, and thus it is more sensitive to the value of $e$ than the other models. 

The violin plots of patient-specific BIC values for the (B) LK, (D) NS, and (F) Emax models show that NS and Emax are better supported by the clinical dataset than LK, see Figure~\ref{fig:violinplot}. The median and mean BIC values for each model are $(2082.13, 16498.42)$ for LK, $(476.21, 4310.81)$ for NS, and $(475.39, 4155.43)$ for Emax. Thus, by BIC, this analysis suggests that the NS and Emax models are comparably suitable for this dataset over the LK model. 


\begin{figure}[htbp!]
    \centering
    \includegraphics[width=1\linewidth]{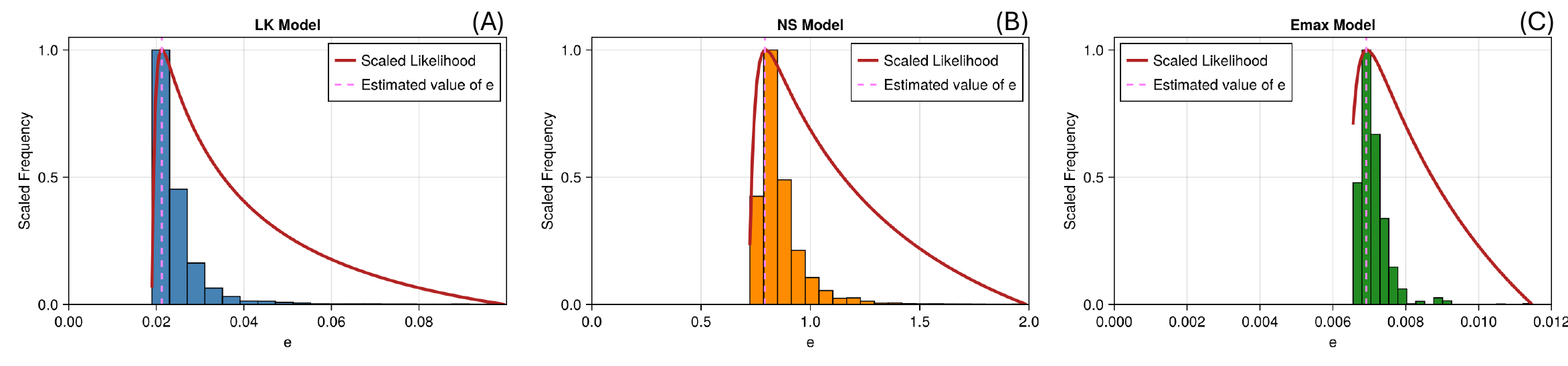}
    \caption{Posterior distributions of CPA efficacy ($e$) estimated at the cohort level for the three chemotherapy models: LK (A), NS (B), and Emax (C).}
    \label{fig:likelihood}
\end{figure}

Figure~\ref{fig:likelihood} displays the posterior distributions and maximum likelihood estimates (MLEs) of the inferred drug efficacy at the cohort level (where all patients were fitted simultaneously). The estimated CPA efficacy was $0.0212$ for the LK model with a $95\%$ credible interval (CI) of $(0.0200, 0.0420)$, $0.809$ for the NS model with a CI of $(0.804, 1.200)$, and $0.00692$ for the Emax model with a CI of $(0.00667, 0.00795)$. We note that these MLEs are all larger than the corresponding individual estimates reported in Figure~\ref{fig:violinplot}. As seen in the individual fitting example above, all three posterior distributions have a relatively tight uni-modal shape, but their supports differ substantially. The narrow posteriors suggest the data is sufficient to inform the parameter's posterior, indicating practical identifiability.

Taken together, these analyses can help inform a more complete model selection rather than relying solely on the information criterion. For example, the BIC analysis suggests the NS and Emax models are the best choice for the clinical dataset. The optimized Emax model delivers larger doses early on while the NS model delivers equal doses. The sensitivity analysis shows that these models are both fairly sensitive to the drug efficacy parameter, and the NS model estimates of patient-specific drug efficacy result in a wide range of values while the Emax model estimates are comparably compact. The LK model, on the other hand, is the simplest model but the BIC scores are the highest suggesting a worse overall fit with the dataset. The LK model is less sensitive to drug efficacy than the other models, with a medium support of estimated values, and an optimized drug delivery schedule that back-loads doses on the last possible days. Thus, one could choose the Emax model due to its low BIC scores, high sensitivity to drug efficacy, and compact support of parameter estimates. Or, one could choose the LK model for it's relative simplicity and lack of sensitivity to the drug efficacy parameter. Or, one could chose the NS model because its bias most closely reflects clinical practice of equalized doses and it also has low BIC scores. Ultimately, the choice should be informed by these types of analyses and by the motivating questions and purpose of the model.

\section{Discussion}

This study demonstrates how the assumptions underlying the functional forms used in  standard cancer treatment mathematical models can significantly affect predicted outcomes. By comparing three different models of chemotherapy and radiation therapy, we showed how the bias of each model is reflected in optimal treatment scheduling, as well as in the sensitivity and resulting inferred parameter variability of these models.

For chemotherapy, we found that the LK chemotherapy model biases optimized schedules towards administration of larger doses on late treatment days, whereas the Emax model biases optimized schedules towards larger doses on early treatment days. Interestingly, the NS model did not display such a bias, as the model predicted equally effective treatment regardless of the tested schedule or dose distribution.

Similarly, for radiotherapy, we found that the LQ model biases optimized schedules towards administration of the largest doses on the latest treatment days, while the PSI model biases schedules towards largest doses on the earliest treatment days. The CDR model predicted optimal tumour reduction when doses were approximately equally fractionated, but did bias optimal schedules towards the latest treatment days. The fact that the CDR model is less sensitive to dose distribution than the LQ and PSI models, and that it predicts maximal tumour reduction for equal dose fractionation (clinical practice) is an advantage of this model.

Next, we paired the chemotherapy and radiotherapy models together, matching their perceived bias. Our results demonstrated that model forms significantly impact  optimized treatment schedules. While slight improvements are observed for the LK + LQ and NS + CDR model combinations with concurrent schedules compared to sequential ones, the optimal concurrent schedules for these pairs contradict those of the Emax + PSI model pair. The Emax + PSI pair predicts greater treatment efficacy compared to the other pairs, but it is also highly sensitive to the applied concurrent protocol. In contrast, the NS + CDR pair shows the most robustness to the applied protocol.

Simulations of adaptive therapy were used to demonstrate the potential effect of model bias in simulation-based treatment planning and prediction. 
The number of treatment cycles and thus overall delivered dose was found to be dependent on both model selection and dose distribution. When sensitive and resistant cells were considered, the time to treatment failure varies by choice of model and dose distribution for both the LK + LQ and Emax + PSI model pairs.


Finally, the clinical parameterization of the prostate cancer dataset provided a concrete example of the impact of model bias, sensitivity, and practical parameter identifiability on model selection. Patient-specific drug efficacy estimates show broad variability in both the estimated best-fitting parameter value and the resulting BIC score. For example, while the LK model is arguably the simplest chemotherapy model, it did not fit well with the given dataset (high BIC scores) despite having a reasonably compact support for the efficacy estimates with a correspondingly low global sensitivity index. 

This highlights the need for thorough analyses and understanding of model bias, sensitivity, and practical identifiability and examination of the inferred parameter posteriors, when building mathematical models for clinical decision making. All of which should be included as a part of the overall uncertainty quantification analysis, and as a part of the model selection process, rather than relying on metrics like the BIC. 
Optimized protocols and parameter estimates will always reflect the structure of the chosen model. One must then acknowledge this limitation, or account for it, by examining multiple models, applying selection criteria based on available data and analyses, and designing robust protocols that acknowledge uncertainty.

\section{Acknowledgements}
KPW would like to acknowledge and thank Dr. Siv Sivaloganathan for all his guidance, mentorship, and laughs that he shared over the years.

\section{Author Contributions}
Conceptualization (CO and KPW), data curation (CO), formal analysis (CO and KPW), funding acquisition (KPW), investigation (CO and KPW), methodology (KPW), project administration (KPW), supervision (KPW), validation (CO and KPW), visualization (CO and KPW), writing draft, review, and editing (CO and KPW). 

\section{Funding}
This work was supported in part by a Discovery Grant 2018-04205 (KPW) from the Natural Sciences and Engineering Research Council of Canada (www.nserc-crsng.gc.ca), and by the Faculty of Science at Toronto Metropolitan University (CO).



%
%
%

\bibliographystyle{unsrt} 
\bibliography{references}

\end{document}